\documentclass[manuscript, acmsmall, screen]{acmart}
\renewcommand\footnotetextcopyrightpermission[1]{} 
\usepackage{tabularx}  
\usepackage{multirow}
\usepackage{lscape}
\usepackage{tcolorbox}
\usepackage{caption}
\usepackage{subcaption}
\usepackage{adjustbox}
\usepackage{makecell}
\usepackage[ruled,vlined,linesnumbered]{algorithm2e}
\usepackage{enumitem}
\usepackage{colortbl}%

\setlist[itemize]{leftmargin=*}

\newcommand{\compactline}{\looseness=-1}
\newcommand{\grayrow}{\rowcolor[gray]{0.89}}
\newcommand{\graycell}{\cellcolor[gray]{0.89}}
\newcommand{\lightgraycell}{\cellcolor[gray]{0.95}}

\begin{document}

\title{Boosting Pointer Analysis With LLM-Enhanced Allocation Function Detection}

\author{Baijun Cheng}
\affiliation{
	\institution{Peking University}
    \city{Beijing}
	\country{China}
 }
\email{prophecheng@stu.pku.edu.cn}

\author{Kailong Wang}
\authornote{Corresponding authors}
\affiliation{
	\institution{Huazhong University of Science and Technology}
    \city{Wuhan}
	\country{China}
 }
\email{wangkl@hust.edu.cn}

\author{Ling Shi}
\affiliation{%
  \institution{Nanyang Technological University}
  \city{Singapore}
  \country{Singapore}
}
\email{ling.shi@ntu.edu.sg}

\author{Haoyu Wang}
\affiliation{
	\institution{Huazhong University of Science and Technology}
    \city{Wuhan}
	\country{China}
 }
\email{haoyuwang@hust.edu.cn}

\author{Peng Di}
\affiliation{%
  \institution{Ant Group}
  \city{Hangzhou}
  \country{China}
}
\affiliation{
	\institution{University of New South Wales}
    \city{Sydney}
	\country{Australia}
 }
\email{dipeng.dp@antgroup.com}

\author{Ding Li}
\affiliation{%
  \institution{Peking University}
  \city{Beijing}
  \country{China}
  }
\email{ding_li@pku.edu.cn}

\author{Xiangqun Chen}
\affiliation{%
  \institution{Peking University}
  \city{Beijing}
  \country{China}
  }
\email{cherry@sei.pku.edu.cn}

\author{Yao Guo}
\authornotemark[1]
\affiliation{%
  \institution{Peking University}
  \city{Beijing}
  \country{China}
  }
\email{yaoguo@pku.edu.cn}


\begin{abstract}
Pointer analysis is foundational for many static analysis tasks, yet its effectiveness is often hindered by imprecise modeling of heap allocations, particularly in C/C++ programs where custom allocation functions (CAFs) are pervasive. 
Existing approaches largely overlook these custom allocators, leading to coarse aliasing and low analysis precision. 
In this paper, we present CAFD, a novel and lightweight technique that enhances pointer analysis by automatically detecting side-effect-free custom allocation functions. 
CAFD employs a hybrid approach: it uses value-flow analysis to detect straightforward wrappers and leverages Large Language Models (LLMs) to reason about more complex allocation patterns with side effects, ensuring that only side-effect-free functions are modeled as allocators.
This targeted enhancement enables precise modeling of heap objects at each call site, achieving context-sensitivity-like benefits without significant overhead.

We evaluated CAFD on 17 real-world C projects, identifying over 700 CAFs. 
Integrating CAFD into a baseline pointer analysis yields a 38× increase in modeled heap objects and a 41.5\% reduction in alias set sizes, with only 1.4x runtime overhead. 
Furthermore, the LLM-enhanced pointer analysis improves indirect call resolution and discovers 29 previously undetected memory bugs, including 6 from real-world industrial applications. 
These results demonstrate that precise modeling of CAFs has the capability to offer a scalable and practical path to improve pointer analysis in large software systems.

\end{abstract}

\maketitle

 \vspace{-3mm}
\section{Introduction}
Pointer analysis is a cornerstone of modern static program analysis, providing essential information about memory reference relationships between program variables. This fundamental technique enables a wide range of applications including bug detection, compiler optimization, and program understanding. In particular, pointer analysis serves as the foundation for advanced analyses such as value flow tracking~\cite{PSVFA, SVF, pinpoint, Falcon}, type state analysis~\cite{ESP, PSTA}, and abstract interpretation~\cite{ikos,Infer}. 
The precision of pointer analysis directly impacts the effectiveness of many downstream applications such as bug detection and compiler optimization~\cite{das2001estimating}, making it a critical component of the software engineering toolchain.

Despite decades of research efforts, existing pointer analysis approaches still face significant limitations that restrict their practical applicability. The most prominent challenge is the fundamental trade-off between precision and scalability. 
For example, context-sensitive pointer analyses can achieve high precision by distinguishing different calling contexts, but they often incur prohibitive computational costs. 
Even a modest k-callsite-sensitive analysis with k=2 can result in a hundredfold overhead compared to context-insensitive analysis~\cite{PUS,Cutshortcut,Graphick}.
Recent optimization techniques such as selective context sensitivity~\cite{Zipper, Scaler, Graphick} and demand-driven analysis~\cite{SUPA, SUPA_extend, DDPA, boomerang} have partially mitigated this issue, but the computational burden remains substantial for large-scale software systems.

\begin{table}[h]
\centering
\caption{Custom and standard AF~(SAFs) counts across projects.} \vspace{-3mm}
\resizebox{0.35\textwidth}{!}{%
\begin{tabular}{c|ccccc}
\toprule
project & bash & git & nasm & openssl & vim \\
\midrule
CAF & 755 & 384 & 118 & 435 & 546 \\
SAF & 42 & 8 & 3 & 3 & 3 \\
\bottomrule
\end{tabular}\vspace{-5mm}
\label{tab:AF_count}
}
\end{table}

\begin{figure}[t]
  \centering  \includegraphics[width=\textwidth]{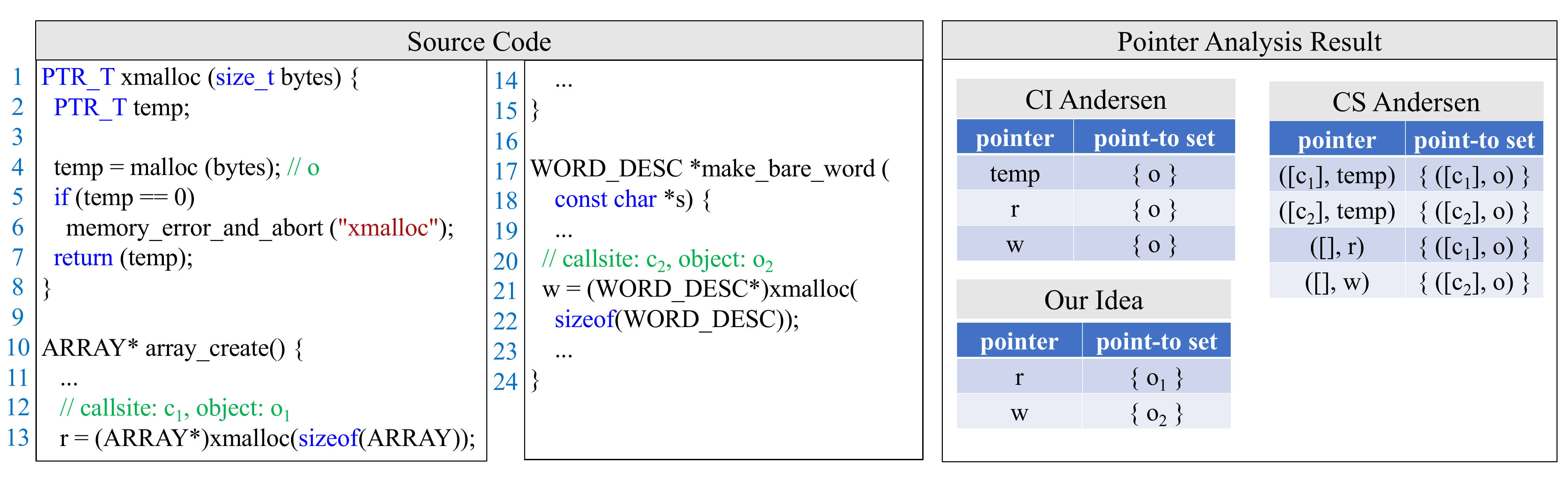}
  \vspace{-3mm}
  \caption{Examples of allocation wrapper from bash. \textbf{CI} denots ``context-insensitive''~(base Andersen), \textbf{CS} denotes ``context-sensitive''.} 
  \label{fig:example}
\end{figure}

A particularly overlooked limitation in current pointer analysis frameworks is their inadequate handling of custom allocation functions~(CAFs). 
Unlike languages such as Java where heap allocation follows a uniform pattern through the new operator, C/C++ programs extensively employ custom allocation wrappers that encapsulate error handling, initialization, and logging logic around standard allocators like \texttt{malloc}. 
These CAFs are not merely occasional utilities, as prior research~\cite {SinkFinder, Goshawk, Raisin} shows that such functions are used extremely frequently in real-world C projects, and accurately modeling them is crucial for detecting memory bugs.
Table~\ref{tab:AF_count} illustrates this by showing the number of calls to the standard AF \texttt{malloc} and one of its custom wrappers in the five real-world projects studied in this work. 
For example, in the bash project as shown in Table~\ref{tab:AF_count}, the custom wrapper \texttt{xmalloc} has 755 call sites compared to only 42 direct \texttt{malloc} calls. 
Current pointer analysis tools typically model only standard allocation functions, treating all objects allocated through custom wrappers as a single abstract object. 
This coarse-grained modeling introduces substantial imprecision, as pointers returned from different call sites of the same wrapper function are incorrectly considered aliases.

CAFs are too widespread in real-world codebases for manual identification and annotation to be feasible~\cite{SinkFinder, Goshawk, Raisin}.
However, identifying CAFs presents two fundamental challenges.
First, the objects they return may already be aliased to other program locations, a behavior that is typically ignored by existing approaches designed to improve heap-related bug detection through custom allocator modeling~\cite{KMeld, SinkFinder, Raisin, PairMiner, PFMiner, Goshawk}.
Second, since CAF identification serves as a pre-analysis step preceding pointer analysis, it must remain computationally lightweight. 
However, precisely analyzing side-effects typically requires expensive pointer analysis, creating a critical tension between accuracy and scalability. These challenges require an efficient strategy that can capture the essential allocation semantics while avoiding the computational burden of full program analysis.

\noindent\textbf{Our Work.} In this paper, we present CAFD (Custom Allocation Function Detector), an automated approach for effectively identifying CAFs. 
Unlike prior methods that uniformly approximate all CAFs as \texttt{malloc}, CAFD is motivated by a pilot study~(\S~\ref{sec:study}) showing that selectively modeling \textit{side-effect free CAFs} (denoted as \textbf{S-CAFs} hereafter, those without caller visible side-effects) can significantly enhance the precision of pointer analysis. 
Using this insight, CAFD focuses on accurately identifying S-CAFs while avoiding misclassification of complex CAFs (\textbf{C-CAFs}) that may introduce side effects and thus cannot be safely treated as \texttt{malloc}.

CAFD is designed as a lightweight pre-analysis step before performing pointer analysis. 
It combines simple \textbf{value-flow tracking} to identify functions where allocation results propagate to return values, with \textbf{LLM assistance} to handle cases involving potential side effects, distinguishing ignorable effects (e.g., in error-handling paths) from those requiring exclusion.
This hybrid design allows CAFD to effectively capture both straightforward and non-trivial allocation patterns with high precision and low cost.

Our experimental evaluation on 17 real-world C projects demonstrates the effectiveness of our approach. 
CAFD successfully identified over 700 CAFs across these projects. 
When integrated into pointer analysis, these identified CAFs led to a 38× increase in the number of modeled heap objects and a 41\% reduction in alias set sizes for allocation-returned pointers, with only 1.4× runtime overhead. 
We further validated the practical impact through two applications: \textbf{(1) indirect call resolution}, where the enhanced analysis reduced candidate sets for 284 indirect calls, and \textbf{(2) memory bug detection}, where it enabled the discovery of 29 additional true bugs that were missed by the baseline analysis, including 6 from Ant Group’s internal industrial application; notably, 23 of these issues have already been fixed in the latest versions or confirmed by developers. 
These results confirm that enhanced allocation function modeling represents a cost-effective approach to improving pointer analysis precision with immediate practical benefits.

\noindent\textbf{Contributions.} Our contributions are summarized as follows:

\begin{itemize}
\item  \textbf{Automated and lightweight identification of CAFs.}
We develop CAFD to automatically identify and model CAFs, enabling finer-grained heap object modeling that reduces imprecision from treating all CAF callsites as aliases.

\item  \textbf{Precision gains with low overhead.}
Our approach achieves precision comparable to full context-sensitive analysis—38× more modeled heap objects and 41\% smaller alias sets—while incurring only 1.4× runtime overhead.

\item  \textbf{Improved real-world analysis results.}
We demonstrate practical impact through better indirect call resolution and memory bug detection, uncovering 30 additional bugs and refining 284 indirect call targets across 17 real-world C projects.

\item \textbf{Tool availability.}
To support reproducibility, we have open-sourced both CAFD~\cite{AFD} and the enhanced pointer-analysis framework~\cite{SEPA}.

\end{itemize}
 \vspace{-3mm}
\section{Background}


\subsection{Pointer Analysis}

\noindent\textbf{Basic Pointer Analysis.} Andersen’s~\cite{Andersen} and Steensgaard’s~\cite{steensgaard} algorithms are two classical flow- and context-insensitive pointer analysis techniques, differing in precision and scalability. Andersen’s analysis uses subset constraints, achieving higher precision with an inclusion-based solution of $O(n^3)$ complexity. In contrast, Steensgaard’s algorithm adopts unification constraints, offering near-linear $O(n)$ performance by merging pointer equivalence classes, at the expense of precision.
Due to its higher precision, Andersen’s analysis is often preferred in scenarios where accuracy is critical~\cite{Falcon, pinpoint, Saber, SUPA, SUPA_extend}. 
In contrast, Steensgaard’s algorithm is better suited for scenarios that demand fast, coarse-grained pointer analysis~\cite{codeql, Fastcheck}.
Table~\ref{tab:andersen_rules} summarizes Andersen’s constraint rules. 
For example, in Figure~\ref{fig:example}, \texttt{xmalloc} allocates a heap object $o$ to \texttt{temp}~(Rule \textbf{Addr}), which is then assigned to \texttt{r}~(Rule \textbf{Copy}) in \texttt{array\_create} and to \texttt{w} in \texttt{make\_bare\_word}. 
This causes (1) spurious aliasing between \texttt{r} and \texttt{w}, since both point to $o$, and (2) potential false negatives in bug detection, e.g., when one allocation from \texttt{xmalloc} is freed while another is not, a memory leak may occur, but Andersen-based detectors may miss it.

\noindent\textbf{Context-sensitive Pointer Analysis.} To overcome the precision limitations of base Andersen analysis, context-sensitive analysis distinguishes objects and pointers based on their calling contexts. Each function call is tagged with a unique context identifier.
For example, in Figure~\ref{fig:example}, context-sensitive analysis differentiates the \texttt{malloc} call in \texttt{xmalloc} by creating two context-tagged objects: $([c_1], o)$ for \texttt{r} in \texttt{array\_create} and $([c_2], o)$ for \texttt{w} in \texttt{make\_bare\_word}. 
This avoids spurious aliasing between \texttt{r} and \texttt{w}, enabling more accurate, context-aware bug detection.
Although context sensitivity greatly enhances precision, it comes at a high computational cost. 
Previous studies~\cite{PUS, Cutshortcut, Graphick} show that, in C programs, 1-callsite sensitivity can cause a 10–15× slowdown, while in Java, 2-object sensitivity may increase analysis time by over 100× compared to context-insensitive analysis.
Although various optimizations~\cite{Zipper, Scaler, Graphick, PUS, DEA, WP_DP, SUPA, SUPA_extend} can reduce overhead, the inherent complexity gap between context-sensitive and -insensitive analyses remains, presenting a fundamental tradeoff between precision and scalability.

\begin{table}[htbp]
\centering
\caption{Rules of Andersen-style pointer analysis.}
\vspace{-3mm}
\label{tab:andersen_rules}
{\small
\begin{tabularx}{\textwidth}{c|c|c|c|c|c}
\hline
\textbf{Type} & \textbf{Statement} & \textbf{Constraint} & \textbf{Type} & \textbf{Statement} & \textbf{Constraint} \\ 
\hline
Addr   & $p = \&o$       & $o \in \text{pts}(p)$ &
Copy   & $p = q$         & $\text{pts}(q) \subseteq \text{pts}(p)$ \\
Store  & $*p = q$        & $\forall o \in \text{pts}(p),\; \text{pts}(q) \subseteq \text{pts}(o)$ &
Load   & $p = *q$        & $\forall o \in \text{pts}(q),\; \text{pts}(o) \subseteq \text{pts}(p)$ \\
Field & $p = q.f$  & $\forall o \in \text{pts}(q),\; o.f \in \text{pts}(p)$ &
Phi & $p = \phi(r, q)$ & $\text{pts}(r) \cup \text{pts}(q) \subseteq \text{pts}(p)$  \\
\hline
\end{tabularx}\vspace{-3mm}
}
\end{table}

\subsection{Allocator Identification}

A promising way to break the precision–scalability trade-off in pointer analysis is to automatically identify custom allocators. 
For example, in Figure~\ref{fig:example}, recognizing \texttt{xmalloc} as a CAF allows the basic Andersen algorithm to distinguish \texttt{r} and \texttt{w} as pointing to different objects, thereby improving precision. Since the implementation of CAFs varies widely across projects, manual annotation is impractical. 
To address this, prior studies~\cite{RRFinder, InferROI, SinkFinder, Raisin, KMeld, PairMiner, PFMiner, Goshawk} have proposed automated CAF recognition to enhance bug detectors such as clang static analyzer~\cite{CSA} for heap-related bugs like memory leaks.

Although effective for lightweight bug detection, these methods adopt aggressive heuristics that overlook pointer relations, resulting in unsoundness for pointer analysis. 
To clarify the scope of their modeling capabilities, we introduce two fundamental concepts used throughout this paper: \textbf{side-effect–free CAFs}~(\textbf{S-CAFs}) and complex CAFs~(\textbf{C-CAFs}).

\begin{definition}[S-CAFs: Side-effect–free CAFs]
A function $f$ is a \emph{side-effect–free CAF} (S-CAF) if every execution in which $f$ returns an allocated object performs no externally observable side effects. 
Formally, $f$ does not write pointer data to the memory space pointed by global variables or caller parameters, or return value. 
All mutations inside $f$ (if any) are confined to local variables or freshly allocated objects that do not escape before being returned.
\end{definition}

\begin{definition}[C-CAFs: Complex CAFs]
A function $f$ is a \emph{complex CAF} (C-CAF) if it allocates or returns an object while performing operations that may produce externally observable effects or introduce non-trivial interprocedural dependencies. 
Examples include updating point-to relations of global variables, mutating shared data structures, invoking stateful library functions, or creating aliases whose effects propagate beyond $f$'s scope. 
Such behaviors make C-CAFs significantly harder to precisely classify or analyze by existing pointer analysis frameworks.
\end{definition}

\begin{figure}[t]
  \centering  \includegraphics[width=\textwidth]{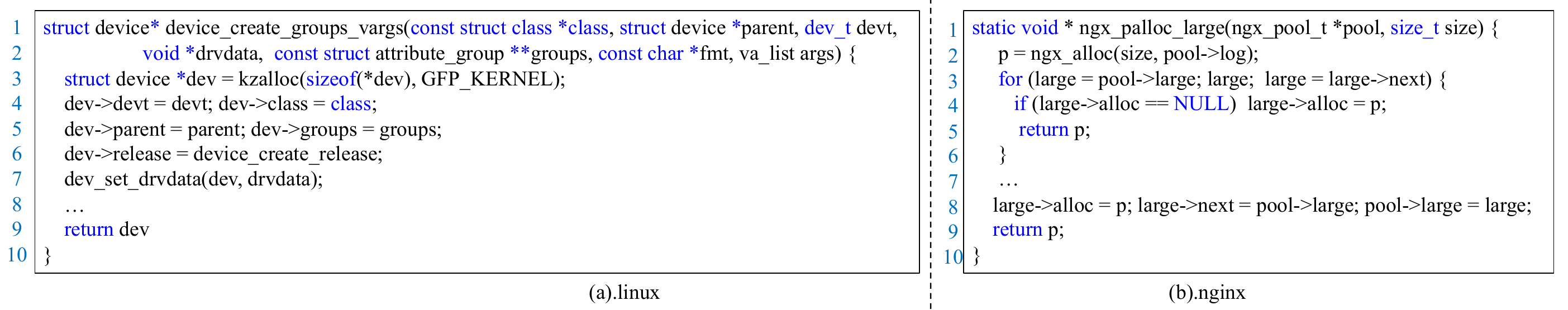}
  \vspace{-4mm}
  \caption{C-CAFs from linux and nginx.}
  \label{fig:complex_example}
  \vspace{-5mm}
\end{figure}

As shown in Figure~\ref{fig:complex_example}, \texttt{device\_create\_groups\_vargs} not only allocates memory but also initializes pointer fields (e.g., function pointer \texttt{release}). 
Ignoring these writes may not affect leak detection, but can cause false positives in null-dereference checks and hinder indirect call resolution. 
Similarly, \texttt{ngx\_palloc\_large} attaches heap memory to a memory pool released in bulk, making it unsuitable to be treated as an S-CAF. 
Hence, existing CAF identification techniques cannot be directly used to enhance pointer analysis.

\section{Pilot Study}~\label{sec:study}
\vspace{-3mm}

To understand the impact of CAF annotation in enhancing the performance of pointer analysis, 
we first conduct a pilot empirical study by manually labeling prevalent S-CAFs in real-world C benchmarks with quantitative evaluations.

\vspace{-2mm}
\subsection{Experimental Setup}~\label{subsec:setup}\vspace{-2mm}

We select eight real-world projects—bash, htop, nanomq, nasm, perl, teeworlds, tmux, and vim—for evaluation. Each project is compiled using LLVM~\cite{llvm}~(version 15.0.0), and its intermediate representation (IR) is extracted from the resulting binaries using wllvm~\cite{wllvm}. The experiments are conducted using SVF~\cite{SVF}~(version 3.0), a widely used framework for C program pointer analysis.
For all projects, we spent approximately eight person-hours reading the source code to identify memory-allocation–related functions. 
For each project, we manually annotated a set of S-CAFs, ensuring that each annotated CAF can be safely approximated as \texttt{malloc} and does not produce side effects observable by its caller. 
This excludes, for example, functions that store the allocated heap object’s address into a caller-provided pointer parameter or assign its fields to point to other memory regions.
Table~\ref{tab:manual_AF_num} presents the number of S-CAFs manually annotated for each project.
During benchmark compilation, we make the following modifications to the default build configuration:

\begin{itemize}
\item Disabling opaque pointers~\cite{OpaquePointers} to retain type information for downstream analysis tasks.

\item Statically linking project libraries when possible, so that all relevant code is preserved within a single bitcode module.

\item Applying the \texttt{mem2reg} optimization pass~\cite{mem2reg} to eliminate unnecessary store/load instructions.
\end{itemize}

Table~\ref{tab:benchmark} summarizes the detailed benchmark statistics, including source lines of code, the number of LLVM IR instructions in each compiled program. (Note: nine additional projects are used for evaluation in the later experiment.)

\begin{table}[htbp]
\centering
\caption{Benchmark statistics. \textit{TL: total source lines~(kLocs); IN: number of LLVM IR instructions~(K instructions).}}
\vspace{-3mm}
\resizebox{\textwidth}{!}{
\begin{tabular}{c|c|c|c|c|c|c|c|c|c|c|c|c|c|c|c|c|c}
\hline
\textbf{Project} & bash & curl & git & htop & lighttpd & nanomq & nasm & openssl & perl & screen & systemd & teeworlds & tmux & vim & wine & h2o & P1 \\
\hline
\textbf{Version} & 5.2 & 8.14.1 & 2.47.0 & 3.3.0 & 1.4 & 0.22.10 & 2.16.03 & 3.4.0 & 5.40.0 & 5.0,1 & 257.6 & 0.7.5 & 3.5 & 9.1.0857 & 9.22 & 2.2.6 & Nan \\
\hline
\textbf{TL} & 400.5 & 382.6 & 1083.9 & 49.6 & 114.0 & 586.4 & 85.9 & 1140.6 & 1527.1 & 40.1 & 1214.2 & 140.2 & 94.4 & 1386.5 & 5989.0 & 166.8 & 646.3 \\
\hline
\textbf{IN} & 187.2 & 114.8 & 586.1 & 38.3 & 64.1 & 219.8 & 69.6 & 639 & 310 & 85.3 & 3827.5 & 188 & 113.5 & 629.1 & 276.5 & 125.9 & 270.2 \\
\hline
\end{tabular}
}
\label{tab:benchmark}\vspace{-3mm}
\end{table}

\begin{table}[h]
\caption{The number of manually labeled CAFs.}
\vspace{-2mm}
\centering
\resizebox{0.62\textwidth}{!}{%
\begin{tabular}{c|c|c|c|c|c|c|c}
\hline
\textbf{bash} & \textbf{htop} & \textbf{nanomq} & \textbf{nasm} & \textbf{perl} & \textbf{teeworlds} & \textbf{tmux} & \textbf{vim} \\
\hline
29 & 5 & 26 & 26 & 3 & 2 & 23 & 34 \\
\hline
\end{tabular}
}
\label{tab:manual_AF_num}
\vspace{-4mm}
\end{table}

\vspace{-1mm}
\subsection{Evaluation Metrics}~\label{subsec:metrics}\vspace{-2mm}

At present, there is no uniform methodology for evaluating the precision of pointer analysis. 
Prior works~\cite{DDPA, VSFS, PUS, SUPA, pinpoint} on C pointer analysis have largely focused on optimizing the computation process by eliminating unnecessary points-to set propagation to reduce time and memory costs. 
As a result, evaluations in these studies primarily emphasize performance improvements and are sometimes supplemented by vulnerability-detection results, but they do not provide a direct assessment of analysis precision.
While other works~\cite{Zipper, Scaler, Graphick, Cutshortcut} on Java pointer analysis typically rely on techniques such as call-graph construction and type-cast analysis to measure analysis precision. 
This is feasible in Java because every allocated object has a well-defined static type; by identifying the set of objects a pointer may reference and inspecting their types, one can determine whether virtual dispatches or type casts are safe.
In contrast, the C memory model is substantially more primitive. 
Memory allocation functions such as \texttt{malloc} return objects of type \texttt{void*}, and C allows highly permissive type conversions between pointer-type variables. 
As a result, type-cast–based evaluation techniques that are effective for Java are not applicable for assessing the precision of pointer analysis in C.
Furthermore, a prior study~\cite{Cocktail} shows that indirect calls in C often arise through globals (23\%, with 97\% stationary), function-pointer struct fields (74\%, with 61\% through stationary fields), and function parameters (76\%). 
Because many of these targets are determined by externally initialized stationary locations, increased pointer-analysis precision does not necessarily yield more precise call resolution. 
Motivated by this observation, we do not use call-graph analysis as our primary evaluation metric.
Instead, we treat it as an application to be analyzed later.

To address the lack of precision-oriented evaluation in prior works, we propose a methodology for assessing the impact of S-CAF annotations on pointer analysis. 
We first introduce several key notations in Table~\ref{tab:notations}, which underpin our evaluation methodology. 
After annotating S-CAFs, the original allocation sites within S-CAFs are ignored and replaced by a set of modeled objects, while allocation sites not covered by any S-CAF remain in the analysis. 
Accordingly, memory objects are categorized into three sets: the killed set ($K$) for eliminated allocation sites (typically those within annotated S-CAFs), the added set ($A$) for newly modeled objects, and the remaining set ($R$) for uncovered allocation sites that persist. 
Based on the $\text{pbs}$ and $\text{as}$ notations in Table~\ref{tab:notations}, we define their corresponding sets for original and enhanced analyses. 
Specifically, $\text{pbs}_o(o)$ and $\text{pbs}_e(o)$ denote the points-by sets of object $o$ in the original and enhanced analyses, respectively, while $\text{as}_o(p)$ and $\text{as}_e(p)$ represent the alias sets of pointer $p$ in the original and enhanced analyses.
For example, in Figure~\ref{fig:example}, we have $o \in K$ and $o_1, o_2 \in A$. 
The points-by sets are $\text{pbs}_o(o) = \{r, w, \text{temp}\}$, $\text{pbs}_e(o_1) = \{r\}$, and $\text{pbs}_e(o_2) = \{w\}$. 
Correspondingly, the alias sets are $\text{as}_o(r) = \{w, \text{temp}\}$ and $\text{as}_e(r) = \emptyset$.

\begin{table}[t]
\centering
\caption{Notations and their meanings used in this study}
\label{tab:concepts}
\renewcommand{\arraystretch}{1.25}

\begin{adjustbox}{width=\textwidth,center}
\begin{tabular}{c|p{0.45\textwidth}|p{0.5\textwidth}}
\toprule
\textbf{Notation} & \makecell{\textbf{Meaning}} & \makecell{\textbf{Example}} \\
\midrule
$K$ & A set of killed allocation sites~(memory object) that are removed after S-CAF annotation.
& In Figure~\ref{fig:example}, the statement \texttt{temp = malloc(byte)} allocates object $o$. 
After modeling \texttt{xmalloc} as an allocator, all statements inside \texttt{xmalloc} are excluded from analysis, hence $o \in K$. \\
\hline

$A$
& The set of newly added allocation sites (memory objects) that replace those in the killed set $K$.
& In Figure~\ref{fig:example}, 
after modeling \texttt{xmalloc} as an allocator, \texttt{r = xmalloc(...)} and \texttt{w = xmalloc(...)} will be modeled as allocation site, allocating $o_1$, $o_2$ respectively, hence $o_1, o_2 \in A$. \\
\hline

$R$
& The set of retained allocation sites that are not covered by any S-CAF and therefore remain unchanged.
& A \texttt{malloc} call in a non-S-CAF function is preserved in the enhanced pointer analysis. \\
\hline

$\text{ar}$ 
& A map that associates each object $o$ with its allocation receiver pointer, i.e., the pointer that receives the object upon allocation.
& For the statement \text{p = \&o}, we have $\text{ar}(o) = p$.\\
\hline

$\text{pbs}$ 
& A map that associates each object $o$ with the set of top-level pointers that may point to it in the point-to analysis.
& For the statement \text{p = \&o}, we have $p \in \text{pbs}(o)$. \\
\hline

$\text{as}$
& A map that associates each pointer $p$ with its alias set in the point-to analysis, i.e., the set of pointers that may refer to the same object as $p$.
& In Figure~\ref{fig:example}, before modeling \texttt{xmalloc} as an allocator, we have $\text{pbs}(o) = \{\text{temp}, r, w\}$; hence, $\text{as}(r) = \{\text{temp}, w\}$. \\

\bottomrule
\end{tabular}
\end{adjustbox}
\label{tab:notations}
\end{table}

The key insight behind our evaluation is that, before modeling S-CAF, the analysis models relatively few heap allocation sites. 
After introducing S-CAF annotations, more allocation sites can be modeled, which generally reduces the size of points-by sets for each object and produces more precise aliasing information. 
For instance, in Figure~\ref{fig:example}, before enhancement there is a single base allocation site $o \in K$ with $\text{pbs}_o(o) = \{\text{temp}, r, w\}$ and $\text{as}_o(r) = \{w, \text{temp}\}$. 
After enhancement, the S-CAF introduces modeled allocation sites $o_1, o_2 \in A$ corresponding to \texttt{r = xmalloc(...)} and \texttt{w = xmalloc(...)}, resulting in $\text{pbs}_e(o_1) = \{r\}$, $\text{pbs}_e(o_2) = \{w\}$, and no aliasing between r and w. 
The original allocation site \texttt{temp = malloc(...);} inside the S-CAF does not require further pointer analysis. 
Based on this insight, we evaluate the effectiveness of S-CAF annotations from the following three perspectives.

\noindent\textbf{Part 1. Number of modeled heap objects}: A larger number of heap objects typically implies more precise modeling across different contexts, reducing imprecision caused by multiple pointers sharing the same base allocation site. We report: (1) the total heap object count change~(THOC) $(|K \cup R| \rightarrow |A \cup R|)$, and (2) the size of the unchanged part~(SUP) $(|R|)$.

\noindent\textbf{Part 2. Average points-by set size}: (1) For replaced objects, we compute the average points-by set sizes before and after annotation as $k_{\text{mean}}$ and $a_{\text{mean}}$ by equation~\ref{equal:pbs}, we then report $\text{pbs}$ change 1~($\text{PC}_1$): $k_{\text{mean}} \rightarrow a_{\text{mean}}$ and $\text{pbs}$ reduction rate1~($\text{PRR}_1$): $\text{PRR}_1 = 1-\frac{a_{\text{mean}}}{k_{\text{mean}}}$. (2) For remaining objects ($R$ set), we compute the average points-by set size under the original and enhanced pointer analyses as $r_{\text{mean}}^o$ and $r_{\text{mean}}^e$ by equation~\ref{equal:pbs}. 
Then we report $\text{pbs}$ change 2~($\text{PC}_2$): $r_{\text{mean}}^o \rightarrow r_{\text{mean}}^e$ and $\text{pbs}$ reduction rate2~($\text{PRR}_2$): $\text{PRR}_2 = 1-\frac{r_{\text{mean}}^e}{r_{\text{mean}}^o}$.

\noindent\textbf{Part 3. Average alias count change}: For each modeled heap allocation statement of the form $p = \&o$ in enhanced analysis, we compute the average alias count of $p$ before and after enhancement, denoted as $\text{as}^o_{\text{mean}}$ and $\text{as}^e_{\text{mean}}$ through equation~\ref{equal:pbs}, respectively. 
We then report (1) Alias Number Change (ANC): $\text{as}^o_{\text{mean}} \rightarrow \text{as}^e_{\text{mean}}$
and (2) Alias Reduction Rate (ARR): $\text{ARR}(P) = 1-\frac{\text{as}^e_{\text{mean}}(P)}{\text{as}^o_{\text{mean}}(P)}$, the alias relationship reduction rate for a group of pointer $P$, as exhaustively enumerating alias relationships among all pointers incurs prohibitive computational overhead.
For both ANC and ARR, in this pivot study, the pointer group $P$ consists of the allocation receivers to heap objects identified by the enhanced pointer analysis. Formally, $P = \{\text{ar}(o) \mid \forall o \in (A \cup R)\}$.
The motivation behind this definition is as follows. Any pointer that may reference a heap object $o$ must alias with its allocation receiver $p = \&o$. In real programs, however, executions can involve additional S-CAF receivers beyond those explicitly annotated, making an exhaustive enumeration of all such pointers infeasible. 
To mitigate this, we define $P$ as the union of receivers before and after enhancement. 
Because the post-enhancement receiver set subsumes the original one, this yields a conservative yet practical approximation of all relevant pointers for evaluation.

\vspace{-2mm}
{\small
\begin{gather}
k_{\text{mean}} = \frac{\sum\limits_{\forall o \in K} |\text{pbs}_o(o)|}{|K|} 
\quad
r_{\text{mean}}^o = \frac{\sum\limits_{\forall o \in R} |\text{pbs}_o(o)|}{|R|} 
\quad
\text{as}^o_{\text{mean}}(P) = \frac{\sum\limits_{\forall p \in P}|\text{as}_o(p)|}{|P|} 
\quad
\notag
\\
a_{\text{mean}} = \frac{\sum\limits_{\forall o \in A} |\text{pbs}_e(o)|}{|A|}
\quad
r_{\text{mean}}^e = \frac{\sum\limits_{\forall o \in R} |\text{pbs}_e(o)|}{|R|}
\quad
\text{as}^e_{\text{mean}}(P) = \frac{\sum\limits_{\forall p \in P}|\text{as}_e(p)|}{|P|} 
\label{equal:pbs}
\end{gather}
}
\vspace{-2mm}

\noindent \textbf{Example}: In Figure~\ref{fig:example}, we have $o \in K$ and $o_1, o_2 \in A$, so THOC is $1 \rightarrow 2$. Since $R = \emptyset$, we obtain $\text{SUP} = 0$. The sizes of the pointer-base sets are $|\text{pbs}_o(o)| = 3$, $|\text{pbs}_e(o_1)| = 1$, and $|\text{pbs}_e(o_2)| = 1$, yielding $k_{\text{mean}} = 3$ and $a_{\text{mean}} = 1$. Thus, $\text{PC}_1$ is $3 \rightarrow 1$, and $\text{PRR}_1 = 66.7\%$.
Because $R = \emptyset$, both $r^o_{\text{mean}}$ and $r^e_{\text{mean}}$ are NaN, which in turn makes $\text{PC}_2$ and $\text{PRR}_2$ undefined.
The allocation-receiver mapping is $\text{ar} = \{o_1 \rightarrow r,\, o_2 \rightarrow w\}$, so $P = \{r, w\}$. The mean alias set sizes are $\text{as}^o_{\text{mean}} = 2$ (since $r$ aliases with $w$ and temp, and symmetrically for $w$), and $\text{as}^e_{\text{mean}} = 0$ (as the aliases involving $w$ and temp for $r$ are eliminated). Consequently, ANC is $2 \rightarrow 0$, and $\text{ARR} = 100\%$.

\subsection{Evaluation Results}~\label{subsec:emprical_res}

Table~\ref{tab:emprical_result} reports the experimental results of enhancing pointer analysis using these S-CAFs. 
From the THOC perspective, the number of modeled heap objects after enhancement increased significantly, on average 26.6× over the baseline. Although the SUP results show that the size of the $R$ set remains non-negligible (i.e., many objects allocated by the base allocator are retained because they are not wrapped by annotated S-CAFs), which in principle may limit the optimization potential, the average $\text{PRR}_2$ score still reaches 15.4\%. This indicates a meaningful level of indirect refinement.
More importantly, the average $\text{PRR}_1$ and $\text{ARR}$ scores are 49.4\% and 44.2\%, respectively. These results suggest that once certain base allocation sites wrapped by S-CAFs are expanded, their point-by sets shrink substantially. This improvement stems primarily from the reduction of cross-context point-to relations. For example, in Figure~\ref{fig:example}, the scenario where both $r$ and $w$ point to the same object $o$ becomes less frequent after enhancement, thereby improving the precision of the points-to analysis and reducing the number of spurious alias relationships.
Specifically, in vim and bash, the number of modeled heap objects grows from fewer than 100 to over 1,000, accompanied by corresponding improvements in $\text{PRR}_1$, $\text{PRR}_2$, and $\text{ARR}$. In teeworlds, because only a small number of S-CAFs are manually annotated, the number of modeled heap objects changes little; nevertheless, the precision-related metrics ($\text{PRR}_1$, $\text{ARR}$) still show substantial improvement. For perl, although only three S-CAFs are marked, the number of modeled heap objects increases from 47 to 272, largely because many heap allocations in perl are performed through S-CAFs.

\vspace{2mm}

\noindent
\begin{tcolorbox}[size=title, opacityfill=0.1, nobeforeafter]
\textbf{Key Findings:} \textit{Marking S-CAFs reduces undesired aliasing between return-value pointers in higher-level callers, thereby mitigating incorrect alias relations. 
This enhancement improves the precision of pointer analysis while maintaining a context-insensitive setting.}
\compactline
\end{tcolorbox}

\begin{table}[htbp]
\centering
\caption{Empirical evaluation results. \textit{\textbf{THOC}: total heap object change, \textbf{SUP}: size of unchanged part, \textbf{PC}: points-by set~(pbs) size change, \textbf{PRR}: pbs reduction rate, \textbf{ANC}: alias number change, \textbf{ARR}: alias reduction rate.}$\uparrow$ indicates that higher values correspond to better results.}
\vspace{-2mm}
\renewcommand{\arraystretch}{1.2}
\setlength{\tabcolsep}{4pt}
\resizebox{0.85\textwidth}{!}{%
\vspace{-2mm}
\begin{tabular}{c|cc|cccc|cc}
\hline
\multirow{2}{*}{\textbf{Project}}  
& \multicolumn{2}{c|}{\textbf{Part 1}} 
& \multicolumn{4}{c|}{\textbf{Part 2}} 
& \multicolumn{2}{c}{\textbf{Part 3}} \\
\cline{2-9}
& \textbf{THOC} & \textbf{SUP} & \textbf{$\text{PC}_1$} & \textbf{$\text{PRR}_1$}(\%) $\uparrow$ & \textbf{$\text{PC}_2$} & \textbf{$\text{PRR}_2$}(\%) $\uparrow$ & \textbf{ANC} & \textbf{ARR}(\%) $\uparrow$ \\
\hline
\multirow{1}{*}{bash} & 49 $\rightarrow$ 1.04K & 41 & 33.11K $\rightarrow$ 11.81K & 64.3 & 10.27K $\rightarrow$ 6.46K & 36.7 & 32.2K $\rightarrow$ 11.6K & 64.0 \\
\hline

\multirow{1}{*}{htop} & 5 $\rightarrow$ 137 & 2 & 11.04K $\rightarrow$ 77.6K & 29.7 & 5.24K $\rightarrow$ 5.11K & 2.5 & 10.27K $\rightarrow$ 7.37K & 28.3 \\
\hline

\multirow{1}{*}{nanomq} & 284 $\rightarrow$ 1.14K & 274 & 67.99K $\rightarrow$ 44.98K & 33.8 & 27.18K $\rightarrow$ 21.53K & 20.8 & 55.24K $\rightarrow$ 34.82K & 37.0 \\
\hline

\multirow{1}{*}{nasm} & 5 $\rightarrow$ 264 & 1 & 14.6K $\rightarrow$ 3.36K & 77.0 & 13.06K $\rightarrow$ 9.75K & 25.4 & 14.58K $\rightarrow$ 4.09K & 72.0 \\
\hline

\multirow{1}{*}{perl} & 47 $\rightarrow$ 272 & 44 & 92.51K $\rightarrow$ 62.44K & 32.5 & 75.91K $\rightarrow$ 74.66K & 1.6 & 89.72K $\rightarrow$ 64.37K & 28.3 \\
\hline

\multirow{1}{*}{teeworlds} & 82 $\rightarrow$ 142 & 76 & 52.43K $\rightarrow$ 16.42K & 68.7 & 22.52K $\rightarrow$ 22.2K & 1.4 & 40.67K $\rightarrow$ 25.91K & 36.3 \\
\hline

\multirow{1}{*}{tmux} & 17 $\rightarrow$ 703 & 14 & 37.54K $\rightarrow$ 19.37K & 48.4 & 23.39K $\rightarrow$ 20.22K & 13.6 & 30.08K $\rightarrow$ 15.88K & 47.2 \\
\hline
\multirow{1}{*}{vim} & 30 $\rightarrow$ 1.76K & 29 & 184.7K $\rightarrow$ 109.76K & 40.6 & 54.88K $\rightarrow$ 43.35K & 21.0 & 182.66K $\rightarrow$ 109.23K & 40.2 \\
\hline
\end{tabular}
}
\label{tab:emprical_result}
\end{table}

\section{Automated S-CAF Identifcation}

Having confirmed that annotating side-effect-free CAFs~(S-CAFs) improves pointer analysis precision (\S~\ref{subsec:emprical_res}), we propose an automated identification strategy that is restricted to this specific class of functions. 
Unlike prior CAF identification techniques~\cite{KMeld, SinkFinder, Raisin, InferROI}, which typically treat all CAF-like functions uniformly, our approach explicitly excludes C-CAFs to preserve both soundness and precision.

\begin{figure}[t]
  \centering  \includegraphics[width=0.8\textwidth]{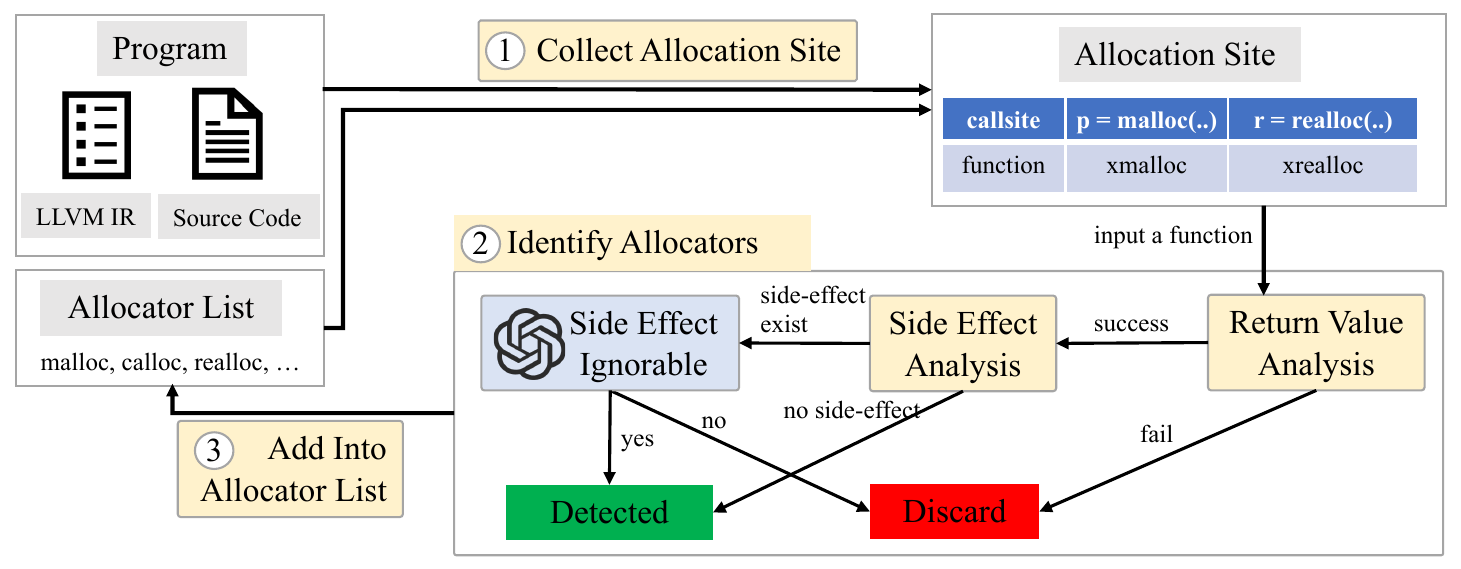}
  \vspace{-4mm}
  \caption{Overview of CAFD.}
  \vspace{-4mm}
  \label{fig:approach}
\end{figure}

\subsection{Design Principles and Rules}~\label{subsec:principle}

In this section, we discuss the key factors considered in the design of our automated strategy for identifying S-CAFs.
Our approach follows a conservative principle to ensure soundness and maintain the precision of pointer analysis. 
Specifically, the strategy is guided by the following general principles~(\textbf{P}) and derivative rules~(\textbf{R}).

\noindent\textbf{P1: Heap-returning behavior only}.
Following SVF~\cite{SVF}, we focus solely on whether a function returns a heap object, ignoring attributes such as allocation size or associated non-pointer data.

\noindent\textbf{P2: Bottom-up propagation}.
Since one S-CAF may internally call another (e.g., in bash, other S-CAFs often wrap \texttt{xmalloc}), S-CAF identification proceeds bottom-up along the call graph. This enables iterative discovery via propagation.

\noindent\textbf{P3: Side-effects}. 
We consider only store operations affecting pointer data, as loads or stores of non-pointer data do not impact the caller. 
Storing \texttt{NULL} is treated as safe, while calls to deallocation functions (e.g., \texttt{free}) are considered side-effecting. 
This design is based on the observation that if an CAF returns a heap object, internal load operations affect only local computation.

\noindent\textbf{P4: Lightweight design}. 
To minimize overhead, we conservatively treat non-trivial store instructions as potential side effects, avoiding full pointer analysis. Error-handling paths are identified efficiently using LLMs rather than complex static analysis. 
The goal is to capture sufficient S-CAFs to enhance pointer analysis precision, without exhaustively enumerating all S-CAFs.

\noindent\textbf{R1: Consistent return values}.
All return values must originate from S-CAF calls or be \texttt{NULL}, ensuring that each execution path returns a pointer to a single heap object. 
Functions returning parameters or stack objects (e.g., Figure~\ref{fig:complex_example} (a)) are excluded.

\noindent\textbf{R2: Immediate return of allocated objects}.
All S-CAF calls must be returned directly to the caller. 
While conservative, this rule aligns with common S-CAF patterns and avoids allocation without ownership transfer, which may lead to memory leaks.

\noindent\textbf{R3: No observable side effects outside error-handling}. 
Prior works~\cite{SVF, pinpoint} assume that system allocator calls return a valid heap object (i.e., they do not model \texttt{malloc} returning \texttt{NULL}). 
User-defined wrappers frequently add diverse logic to handle the situation of allocators returning \texttt{NULL}, but, following this common assumption, those error-handling branches can be ignored in our analysis. 
Our observation of real-world C code further supports this simplification: the normal execution paths of such wrappers typically either return a newly allocated object or immediately initialize only non-pointer fields, whereas their error-handling paths exhibit substantial variability across implementations.

\subsection{CAFD Implementation}~\label{subsec:approach}

Following the principles and rules defined in \S~\ref{subsec:principle}, we design CAFD as illustrated in Figure~\ref{fig:approach}. CAFD takes as input the target program along with an initial set of known system AFs, which initially includes system allocation APIs. It then iteratively analyzes the program to identify additional S-CAFs in accordance with these principles, incrementally adding them to the Allocator List (denoted as \textbf{AL} hereafter). 
The detailed procedure is presented in Algorithm~\ref{alg:getSAFs}. 
Through this process, CAFD discovers new S-CAFs via the following three steps.

\begin{algorithm}[t]
\caption{Identification of S-CAFs}
\label{alg:getSAFs}
\KwIn{Program $program$}
\KwOut{Allocator list $AllocatorList$}
\SetKwProg{Fn}{Function}{}{}

\Fn{\textsc{getSCAFs}$(program)$}{
    $AllocatorList \leftarrow \textsc{getSystemAllocators}()$\;
    $changed \leftarrow true$\;
    \While{$changed$}{
        $callsites \leftarrow \textsc{collectCallSites}(program, AllocatorList)$\;
        $functions \leftarrow \emptyset$\;
        \ForEach{$callsite \in callsites$}{
            $F \leftarrow \textsc{getEnclosingFunction}(callsite)$\;
            \If{$F \notin functions$}{
                $functions \leftarrow functions \cup \{F\}$\;
            }
        }
        $topoFuncs \leftarrow \textsc{topologicalSort}(functions)$\;
        $oldSize \leftarrow |AllocatorList|$\;
        \ForEach{$F \in topoFuncs$}{
          \If{\textsc{IdentifyAllocator}$(F)$}{
                $AllocatorList \leftarrow AllocatorList \cup \{F\}$\;
            }
        }
        \If{$|AllocatorList| = oldSize$}{
            $changed \leftarrow false$\;
        }
    }
    \Return $AllocatorList$\;
}
\vspace{0.5em}

\Fn{\textsc{IdentifyAllocator}$(F)$}{
    \ForEach{$ret \in \textsc{Returns}(F)$}{
        \If{\textbf{not} \textsc{BackWardAnalysis}$(ret)$}{
            \Return false\;
        }
    }
    \ForEach{$call \in \textsc{SCAFCalls}(F)$}{
        \If{\textbf{not} \textsc{ForwardAnalysis}$(call)$}{
            \Return false\;
        }
    }
    $sideEffects \leftarrow \emptyset$\;
    \ForEach{$stmt \in F$}{
        \If{\textsc{isSideEffecting}$(stmt)$}{
            $sideEffects \leftarrow sideEffects \cup \{stmt\}$\;
        }
    }
    \If{$|sideEffects| = 0$}{
        \Return true\;
    }
    $ignorable \leftarrow \textsc{queryLLM}(F, sideEffects)$\;
    \Return $ignorable$\;
}
\end{algorithm}

\noindent\textbf{Step 1: Collect Allocation Callsites.} In this step, we extract a set of functions, denoted as \textsf{CSet}, each of which contains at least one call to a known S-CAF, corresponding to line~5 of Algorithm~\ref{alg:getSAFs}. 
In the algorithm, \textsf{CSet} corresponds to the variable \textsf{functions}. 
Let $P$ denote the analyzed program; this process can be formalized by formula~\ref{equa:CAC}. 
For indirect calls, we leverage Kelp~\cite{Kelp} to resolve their potential targets. 
An indirect call is treated as a S-CAF call only if all of its resolved targets are classified as S-CAFs.  
Since the Allocator List is updated iteratively in the loop, in each round, we do not re-analyze the caller functions of S-CAFs that have already been identified in previous rounds.

\vspace{-2mm}
\begin{equation}
\forall F \in P,\ \left[ \exists\, \text{call } f(r_1, \ldots, r_n) \in F \wedge f \in \text{AL} \right] \Rightarrow F \in \text{CSet}
\label{equa:CAC}
\end{equation}

\noindent\textbf{Step 2: Identify Candidate Functions from Known S-CAF Calls.} 
For each collected callsite (corresponding to the variable \texttt{callsites} in the algorithm~\ref{alg:getSAFs}), we traverse its enclosing function and analyze it in a bottom-up manner along the topological order of the call graph to identify potential new S-CAFs. 
Specifically, for each function, we invoke the \textsc{IdentifyAllocator} procedure to determine whether it qualifies as a S-CAF. 
The \textsc{IdentifyAllocator} procedure consists of three sub-steps~(2.1, 2.2, and 2.3).

\noindent\textbf{Sub-Step 2.1: Return Value Analysis~(Rule R1-R2).} 
The goal of this step is to determine, for a given function $F$ that invokes one or more S-CAFs, whether (1) all S-CAF call receivers are returned to the caller, and (2) all return values are either S-CAF call receivers or \texttt{NULL}.  
To this end, we adopt a simple intra-procedural value-flow tracking strategy that ignores memory operations, inspired by Kelp~\cite{Kelp}.  
For condition (1), we perform forward tracking from each S-CAF call of the form $p = \&o$~(an \textbf{Addr} statement), computing a boolean variable $\text{FW}(p)$ that indicates whether $p$ reaches only the return value (corresponding to lines 23--25 of Algorithm~\ref{alg:getSAFs}).  
For condition (2), we perform backward tracking from each return instruction of the form $\text{return} \; r$, and compute $\text{BT}(r)$, a boolean variable indicating whether $r$ only receives values from S-CAF calls (corresponding to lines 20--22 of Algorithm~\ref{alg:getSAFs}).  
This design is motivated by the following considerations. 
\textbf{First}, both forward and backward tracking of store/load operations require precise pointer analysis, which can be computationally expensive; however, we observe that for many S-CAFs, the value flows from the allocation site to the return value do not involve complex store/load operations. 
\textbf{Second}, in backward analysis, a return value originating from an allocation site indicates that the function successfully returns a heap object, whereas a \texttt{NULL} return indicates allocation failure. 
Returning other values (e.g., directly returning a parameter or an existing object) suggests that under some non-exceptional executions the function may not produce a fresh object, so we conservatively handle such cases by treating the function as not an allocator under all conditions. 
In other words, we conservatively assume it is not S-CAF. 
\textbf{Third}, in forward analysis, if an allocation site does not reach the return value, it may escape via a caller-provided pointer or cause a memory leak, which is difficult to analyze precisely.
Therefore, we conservatively assume such functions cannot be considered S-CAFs.  
The detailed tracking rules are provided in Table~\ref{tab:value_flo_rules}.

\begin{table}[htbp]
\centering
\caption{Rules of forward and backward value flow tracking.}
\vspace{-1mm}
\label{tab:value_flo_rules}
{\small
\begin{tabularx}{0.9\textwidth}{@{}c|c|c|c@{}}
\hline
\textbf{Type} & \textbf{Statement} & \textbf{Forward} & \textbf{Backward} \\ 
\hline
Addr & p = \&o & Nan & \text{BT}(p) = \text{True} \\
\hline

Copy & $p = q$ & $\text{FW}(q) \wedge= \text{FW}(p)$ & $\text{BT}(p) \wedge= \text{BT}(q)$ \\
\hline

Phi & $p = \phi(r, q)$ & $\text{FW}(r) \wedge= \text{FW}(p) \;\;\;  \text{FW}(q)\wedge= \text{FW}(p)$ & $\text{BT}(p) \wedge= \text{BT}(r) \wedge \text{BT}(q)$ \\
\hline

NULL & $p = \text{NULL}$ & Nan & \text{BT}(p) = \text{True} \\
\hline

Store  & $*p = q$ & $\text{FW}(q) = \text{False} \newline \text{FW}(p) = \text{False}$ & $\text{BT}(q) = \text{False} \;\;\; \text{BT}(p) = \text{False}$ \\
\hline

Load   & $p = *q$ & $\text{FW}(q) = \text{False}$ & $\text{BT}(p) = \text{False}$ \\
\hline

Field & $p = \&q\rightarrow f$  & $\text{FW}(q) = \text{False}$ & $\text{BT}(q) = \text{False}$ \\
\hline

Return & $\text{return} \; q$ & $\text{FW}(q) = \text{True}$ & Nan \\
\hline
\end{tabularx}
}
\end{table}

\noindent\textbf{Sub-Step 2.2: Side Effect Analysis.} 
To determine whether a function $F$ contains side effects, we precompute a bottom-up mapping $\text{SI}$~(Side-effect Instructions), which maps each function $F$ to the set of side-effecting statements within it.  
Function $F$ is deemed side-effecting if $\text{SI}[F] \neq \emptyset$.  
The set $\text{SI}[F]$ is constructed as formula~\ref{equa:sidemap}, corresponding to lines 27-29 of Algorithm~\ref{alg:getSAFs}, where $\text{isStoreToNonNullPointer}(s)$ indicates that $s$ is a store of non-null pointer data, and $\text{calls}(s, f)$ denotes that $s$ is a call to function $f$.  
Accordingly, the helper function \textsc{isSideEffecting} returns \texttt{true} for a statement $stmt$ if $stmt \in \text{SI}[F]$.
Since we do not perform a full pointer analysis, this side-effect computation strategy is conservative but efficient, while still remaining sound.

\vspace{-2mm}
\begin{align}
\text{SI}[F] &= \{ s \in F \mid\ 
    \text{isStoreToNonNullPointer}(s)  \lor\ (\exists f.\ \text{calls}(s, f) \land \text{SI}[f] \ne \emptyset) \}
\label{equa:sidemap}
\end{align}

\begin{figure}[t]
  \centering  \includegraphics[width=0.8\textwidth]{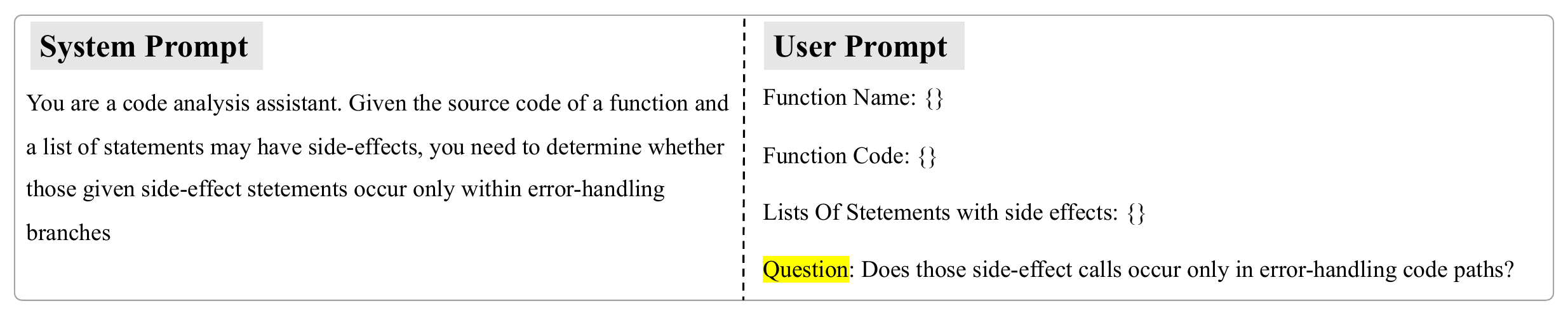}
  \vspace{-4mm}
  \caption{Prompts used for querying LLM.}
  \label{fig:prompt}
  \vspace{-4mm}
\end{figure}

\noindent\textbf{Sub-Step 2.3: Side Effect Ignorable~(Rule R3).} 
This step evaluates whether the side effects of a function $F$ can be safely ignored when $F$ exhibits such effects, corresponding to line 32 of Algorithm~\ref{alg:getSAFs}.
Following SVF~\cite{SVF}, we assume that CAFs correctly allocate heap objects and return pointers, while ignoring non-pointer data and object sizes.
As a result, side effects in the error-handling paths of CAFs that typically return \texttt{NULL} can be disregarded.
However, identifying such error-path side effects using heuristic strategies is challenging, as not all CAFs return \texttt{NULL}, and not all error conditions take the form of checks like \texttt{p == NULL}.
To address this, we use LLMs for lightweight semantic analysis, motivated by recent studies~\cite{GPTScan, RepoAudit, Lara, Latte, LLift, Artemis, IRIS, InferROI, Vul_RAG} that demonstrate their capabilities in code understanding. The task we assign to the LLM is intentionally simple: given a list of side-effecting statements and the code of the target CAF, the LLM is asked to determine whether all such statements reside within error-handling paths. 
Although LLMs may exhibit hallucinations in complex reasoning tasks, the task we assign here is intentionally simple and does not require deep or fine-grained inference. 
As a result, we expect the analysis to remain sufficiently robust despite these limitations.
Accordingly, we employ a straightforward zero-shot prompting approach, with a template illustrated in Figure~\ref{fig:prompt}, where the function name, its source code, and the side-effecting statements are instantiated into the template.

\noindent\textbf{Step 3: Updates and Iteration}. 
Newly identified S-CAFs are added to the AL, and the process repeats until the list is no longer updated (corresponding to line 15 of Algorithm~\ref{alg:getSAFs}).

\begin{figure}[t]
  \centering  \includegraphics[width=0.9\textwidth]{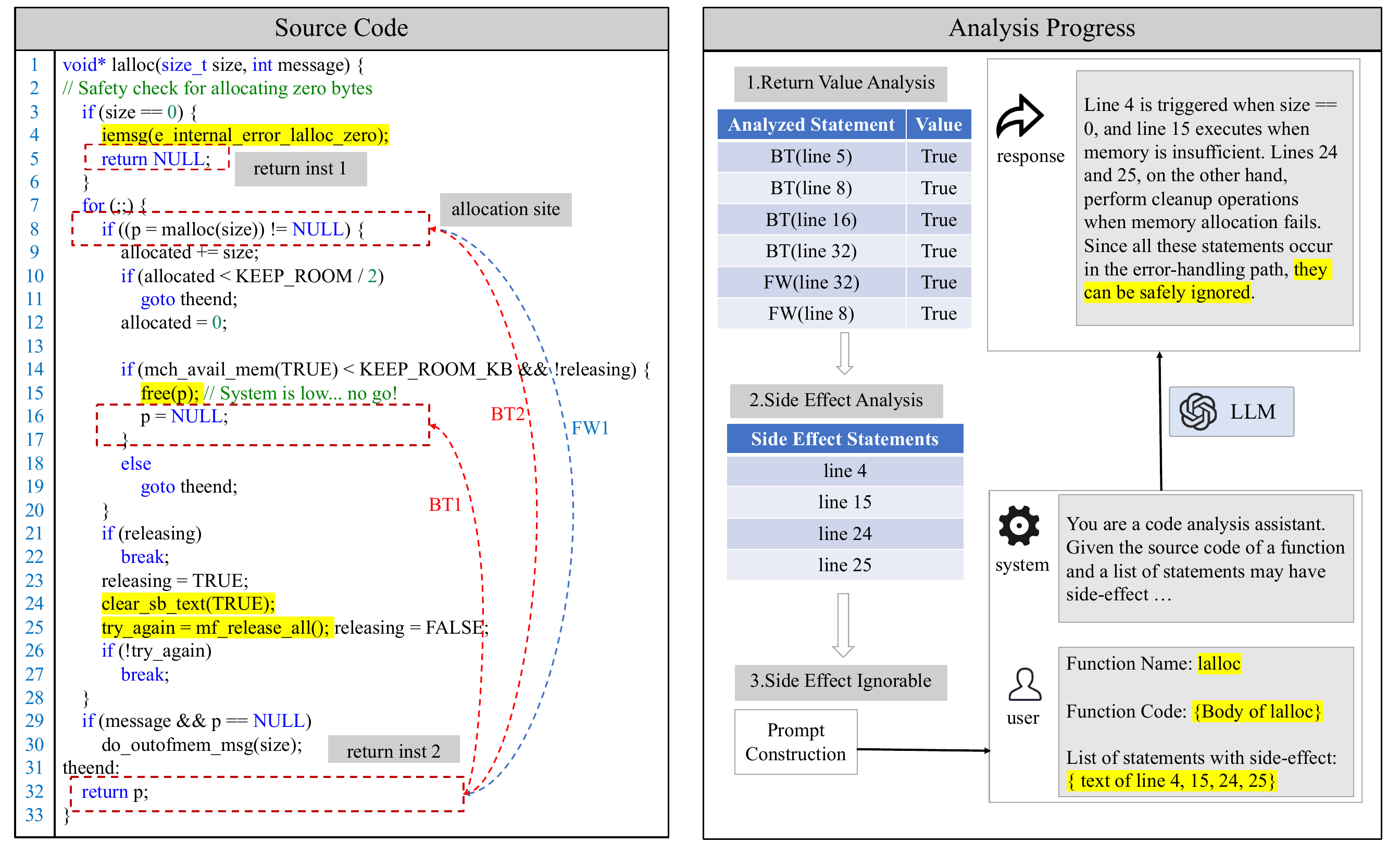}
  \vspace{-4mm}
  \caption{An Illustrative Example of CAFD.}
  \vspace{-4mm}
  \label{fig:running_example}
\end{figure}

\textbf{Illustrative Example}. 
We illustrate the \textsc{IdentifyAllocator} analysis using the function \texttt{lalloc} in Figure~\ref{fig:running_example}. 
\texttt{lalloc} contains two return statements (at line 5 and line 32) and one allocation site (at line 8).  
Applying backward tracking to the return at line 5 shows that its return value is \texttt{NULL}, hence $\text{BT}(\text{line} 5) = \text{True}$. 
The return at line 32 receives values from line 8 (an allocation receiver) and line 16 (\texttt{NULL}); therefore $\text{BT}(\text{line }32)=\text{True}$ as well. 
Together, these results indicate that \texttt{lalloc}'s return value comes only from an allocator or \texttt{NULL}, making it a potential S-CAF.  
For forward tracking, the object allocated at line 8 flows directly to the return at line 32 and is not stored into other variables (e.g., via stores), so $\text{FW}(\text{line} 8)=\text{True}$. 
Consequently, the Return Value Analysis succeeds.  
We then perform side-effect analysis and identify statements at lines 4, 15, 24, and 25 as potentially side-effecting. 
Precisely determining whether these statements affect callers would require expensive pointer analysis.
To avoid this cost, we apply a pragmatic heuristic: we check whether these statements lie on allocation-failure/error-handling paths. 
Using an LLM-based classifier, we conclude that all four statements occur only in allocation-failure branches and thus can be safely ignored. 
With return value and side-effect checks passed, \texttt{lalloc} is accepted as a S-CAF.

\section{Evaluation}~\label{sec:evaluation}

We evaluate the effectiveness of CAFD by answering the following three research questions~(RQs).

\begin{itemize}
\item \textbf{RQ1}: How effective is CAFD in identifying S-CAFs? We evaluate its ability to detect S-CAFs, which are allocator-like functions that introduce no caller-visible side effects and can be soundly approximated as \texttt{malloc}. In contrast, C-CAFs may modify the state of global variables or caller parameters, rendering such an approximation unsound.

\item \textbf{RQ2}: Can the S-CAFs identified by CAFD enhance pointer analysis?
We evaluate whether the S-CAFs discovered by CAFD lead to improved pointer analysis, using the metrics defined in \S~\ref{subsec:metrics}.

\item \textbf{RQ3}: Can CAFD-improved pointer analysis benefit downstream tasks?
We assess whether the enhanced pointer analysis improves the performance of downstream applications, including indirect call analysis and value-flow-based bug detection~\cite{Saber}.
\end{itemize}

\subsection{Experimental Settings}

\textbf{Benchmark Selection}. We select 16 open-source projects and one industrial project from Antgroup~(denoted as \textbf{P1}) as benchmarks, including the 8 used in our pilot study~(\S~\ref{sec:study}) and 9 to substantiate our analysis. 
Detailed information about these programs is provided in Table~\ref{tab:benchmark}. 
Each project is compiled according to the principles in \S~\ref{subsec:setup}.

\noindent\textbf{LLM Selection}. Our tool requires integration with LLMs.
However, using commercial LLMs~\cite{ChatGPT, Claude, gemini} incurs significant token costs and is often prohibitively expensive. 
When analyzing the 16 open-source projects, we follow the setting in~\cite{SEA} and deploy open-source instruction-tuned LLMs~\cite{InstructionTuned}.
Specifically, we select four widely used and recently released models: Qwen3-14B~\cite{Qwen3}, LLaMA-3.2-3B-Instruct~\cite{llama}, Phi-4-Reasoning-Plus~\cite{Phi4}, and DeepSeek-R1-0528-Qwen3-8B~\cite{DeepSeek} (denoted as \textbf{qwen}, \textbf{llama}, \textbf{phi}, and \textbf{ds}, respectively).
For the Ant Group internal project P1, due to confidentiality requirements, all experiments must be executed on company-internal machines. 
However, internal server management policies prevent us from using machines capable of deploying open-source LLMs. 
Consequently, we rely on web-API access to LLMs already deployed in the internal environment.
Due to the limited available resources, we perform the P1 experiments using only the Kimi-K2-Instruct~\cite{kimi-k2} model (denoted as \textbf{kimi-k2} hereafter), and we are unable to use kimi-k2 to analyze the open-source projects.

\noindent\textbf{Implementation}. CAFD is built on top of LLVM~\cite{llvm} 15.0.0, with the value-flow tracking component operating directly on the partially SSA form LLVM IR. 
To support LLM-based analysis, we deploy all models locally using vLLM~\cite{vllm}. 
CAFD queries the locally hosted LLMs via curl~\cite{curl} and retrieves the model outputs.
When querying an LLM, CAFD includes the source code of the function under analysis. 
To enable this, we pre-extract the source code of each function using clang’s libTooling infrastructure~\cite{ClangLibtool} and store it in a database.
For each CAF under analysis, we perform five LLM queries and adopt majority voting to reduce the influence of potential hallucinations.
During IR-level analysis, CAFD retrieves the corresponding source code by matching the function name and line number against this database.
Once S-CAFs are identified, we incorporate them into the base Andersen-style pointer analysis, as implemented in SVF~\cite{SVF}.

\noindent\textbf{Environment}. All experiments were conducted on a machine running Linux 5.15.0-142, equipped with an Intel(R) Xeon(R) Silver 4210R CPU @ 2.40GHz, 128 GB of RAM, and two NVIDIA RTX 3090 GPUs with 24 GB memory each.

\noindent\textbf{Baseline Approaches}. In our evaluation:

\begin{itemize}
\item For RQ1, we compare CAFD with existing allocator detectors KMeld~\cite{KMeld}, Raisin~\cite{Raisin}, and a purely heuristic-based baseline~(denoted as \textbf{CAFD-H}) that removes side effect analysis from CAFD, effectively serving as an ablation.

\item For RQ2, we compare the pointer analysis results under three settings: original pointer analysis, 1-callsite-sensitive analysis~(denoted as \textbf{1ctx} hereafter), CAFD-H enhanced analysis, and CAFD-enhanced analysis.

\item For RQ3, we assess the downstream impact of CAFD by comparing the performance of pointer-analysis-based applications under both the original and CAFD-enhanced analyses. Specifically, we focus on two tasks: indirect call resolution and value-flow-based bug detection.
\end{itemize}

\subsection{RQ1: Effectiveness of CAFD}

In RQ1, we systematically evaluate the performance of CAFD from five perspectives.
To account for potential randomness in LLM reasoning, including hallucinations and other stochastic effects, we configure three temperature settings: 0.4, 0.6, and 0.8. 
These temperatures cover the ranges recommended by the technical reports of the open-source models used in our experiments.
For each model, we run one complete experiment under each temperature.
The results indicate that LLM outputs remain highly stable across these settings.
The models qwen, phi, and kimi-k2 produce identical outputs under all three temperatures. The models ds and llama exhibit differences in only two to three projects. This limited variation arises because the task assigned to the LLMs in CAFD is relatively simple, which reduces the influence of hallucinations or other random factors.
Based on this observation, in the evaluation of the overall effectiveness of CAFD (\$~\ref{subsec:overall_effectiveness}), we report one representative set of results for qwen, phi, and kimi-k2, since the three temperature settings lead to the same outputs. For ds and llama, we report the set of results corresponding to the temperature that performs best.
Using these representative results, we then present the accuracy of CAFD for each LLM (\$~\ref{subsec:acc_afd}), followed by an analysis of the cross-temperature differences (\$~\ref{subsec:difference}). 
We further compare CAFD with baseline approaches using the best-performing LLM (\$~\ref{subsec:AFD_baseline}). 
Finally, we report the token and time cost of each LLM under its representative temperature (\$~\ref{subsec:afd_cost})

\subsubsection{Overall effectiveness of CAFD}~\label{subsec:overall_effectiveness}

To assess the reliability of LLM-assisted decisions, we manually inspect all S-CAFs determined by the LLMs (i.e., those in $\text{num}_2$).
This set is of manageable size. 
In this inspection, we primarily evaluate precision, that is, whether the LLM incorrectly treats a non-ignorable side effect statement as ignorable. Specifically, for each flagged side-effect operation, we read the source code and analyze: (1) whether the operation occurs only within an error-handling path (e.g., cleanup or logging after a failure), and (2) whether it affects pointer states or memory observable by upper-level callers. 
This ensures that no externally visible mutations are mistakenly dismissed. 
The entire process took approximately 3 person-hours. 

For each model, we run three epochs. The results of qwen, phi, and kimi-k2 are identical across all three runs, while llama and ds show only minor differences on two to three projects. We therefore report a representative set of results for qwen, phi, and kimi-k2, and the best-performing epoch for llama and ds.
Table~\ref{tab:afd_res} summarizes the number of S-CAFs identified by CAFD using different strategies (purely heuristic-based vs. LLM-assisted) for each open-source project.
Each entry in the table is presented as ($\text{num}_1$/$\text{num}_2$), where $\text{num}_1$ denotes the total number of detected S-CAFs, and $\text{num}_2$ indicates those determined with the assistance of an LLM.
Under the purely heuristic configuration, every S-CAF detected by CAFD is guaranteed to satisfy the design principles and thus considered sound. 
However, with LLM involvement, S-CAFs may be identified through reasoning provided by the model, which introduces the risk of incorrect classification due to hallucination or misunderstanding. 
Moreover, a misclassified S-CAF may propagate upwards and influence the identification of additional unsafe functions.
After manual analysis, we make the following observations:

\begin{itemize}

\item For qwen, a total of 86 unique S-CAF instances are analyzed. 
We verify that all model predictions are correct in both outcome and reasoning.

\item For llama, 32 S-CAFs are identified. 
Among them, 4 have correct predictions but incorrect or flawed reasoning, and 1 case was both incorrectly reasoned and incorrectly predicted.

\item For phi, 87 S-CAFs are extracted for all projects, and all are manually confirmed to be both sound and correctly reasoned.

\item For ds, 95 S-CAFs are identified, with approximately 14 exhibiting incorrect or inconsistent reasoning and incorrect classification results. 

\item Based on over 80 S-CAFs identified by qwen and phi, CAFD ultimately discovered over 700 S-CAFs across 16 open-source projects. In contrast, a purely heuristic strategy found only 127 S-CAFs. 

\item On the Ant Group project P1, CAFD combined with kimi-k2 identifies a total of 77 S-CAFs, of which 4 are determined with LLM assistance. In contrast, a purely heuristic strategy failed to find any S-CAFs.

\end{itemize}

We notice that the LLM-enhanced strategy identifies substantially more S-CAFs, which is primarily due to its iterative bottom-up approach. 
For instance, consider S-CAF \texttt{FuncA} calling \texttt{FuncB}, and \texttt{FuncB} calling \texttt{FuncC}. 
A purely heuristic strategy can detect \texttt{FuncC}, but fails to identify \texttt{FuncB}. 
By leveraging an LLM to identify \texttt{FuncB}, the bottom-up analysis can then propagate this information to detect \texttt{FuncA}. 
Without the LLM, only \texttt{FuncC} would be discovered. 
The LLM-assisted bottom-up approach thus enables detection of both \texttt{FuncB} and \texttt{FuncA}, substantially increasing overall S-CAF coverage.
These findings suggest that instruction-tuned open-source LLMs can be effectively integrated into CAFD, though their reliability varies. 
qwen and phi exhibit high precision, while ds and llama occasionally introduce hallucinations or misclassifications.
As illustrated in Figure~\ref{fig:af_example}(a), qwen accurately identifies that lines 3, 10, and 13 in the \texttt{lalloc} function, though potentially containing side effects, are part of error-handling logic and can therefore be safely ignored.
Compared to the conservative heuristic-based approach, LLMs more effectively identify S-CAFs by leveraging semantic understanding, enabling CAFD to capture diverse patterns that heuristics often miss.

\begin{table}[htbp]
\caption{Overall statistics of each project for CAFs detected by CAFD.}\vspace{-2mm}
\centering
\renewcommand{\arraystretch}{1.1}
\resizebox{\textwidth}{!}{%
\begin{tabular}{c|cccccccccccccccc|c}
\toprule
\textbf{Project} & bash & curl & git & h2o & htop & lighttpd & nanomq & nasm & openssl & perl & screen & systemd & teeworlds & tmux & vim & wine & \textbf{Total} \\
\midrule
\textbf{heuristic} & 3/0 & 15/0 & 4/0 & 17/0 & 7/0 & 22/0 & 21/0 & 20/0 & 1/0 & 3/0 & 0/0 & 10/0 & 4/0 & 0/0 & 1/0 & 98/0 & 127/0 \\
\hline
\textbf{+qwen} & 65/5 & 16/1 & 54/9 & 17/0 & 10/2 & 22/0 & 56/5 & 20/0 & 139/23 & 11/5 & 4/4 & 24/12 & 4/0 & 96/6 & 62/7 & 115/7 & 715/86 \\
\hline
\textbf{+llama} & 64/4 & 15/0 & 28/3 & 17/0 & 7/0 & 22/0 & 53/2 & 20/0 & 1/0 & 4/1 & 3/3 & 23/10 & 4/0 & 96/6 & 1/0 & 105/3 & 463/32 \\
\hline
\textbf{+phi} & 64/4 & 16/1 & 55/10 & 17/0 & 10/2 & 22/0 & 56/5 & 23/1 & 138/22 & 10/4 & 4/4 & 28/15 & 4/0 & 96/6 & 64/9 & 121/8 & 728/91 \\
\hline
\textbf{+ds} & 65/5 & 16/1 & 54/9 & 17/0 & 10/2 & 22/0 & 56/5 & 23/1 & 150/34 & 10/4 & 4/4 & 21/9 & 4/0 & 97/7 & 63/8 & 114/6 & 726/95 \\
\bottomrule
\end{tabular}
}
\vspace{-3mm}
\label{tab:afd_res}
\end{table}

\subsubsection{Accuracy of CAFD}~\label{subsec:acc_afd} 

We evaluate the effectiveness of the CAFD strategy in identifying S-CAF functions, using precision and recall as the primary metrics. 
Since no ground-truth labels for S-CAF functions are available, we manually audit all functions analyzed by the LLMs and classify them into four categories: true positive (TP), false positive (FP), false negative (FN), and true negative (TN). 
According to the previously defined criteria, if a function is indeed an S-CAF and the LLM correctly recognizes it as such, it is labeled as TP, and the other categories are defined accordingly.
Table~\ref{tab:afd_acc} summarizes the overall performance of the evaluated LLMs, where qwen, phi, llama, and ds represent aggregated results across 16 open-source projects, while kimi-k2 corresponds to a separate evaluation on Ant Group’s internal project P1. 
The results show that all LLMs except ds achieve 100\% precision; the false positives of ds have been analyzed earlier. 
However, all models still exhibit limited recall. The reasons are as follows:
\textbf{1.Intrinsic limitations of LLMs}. Some side-effect statements appear only within error-handling code, but the LLMs incorrectly classify such functions as C-CAF. This issue is relatively minor, accounting for approximately four cases in both qwen and phi.
\textbf{2.Overly conservative analysis strategy}. This accounts for the majority of false negatives and can be further divided into two types. The first involves local side-effect operations that are later cleaned up, contributing roughly 15 and 14 FN cases in qwen and phi, respectively. 
The second involves side effects that do not influence upper-level callees—for instance, modifications to certain state variables unrelated to any heap pointer—which account for around 10 FN cases in both qwen and phi.
Nevertheless, this conservative strategy helps preserve the soundness of subsequent pointer analysis. More advanced agentic approaches would be required to further identify such S-CAF functions.

It should be noted that the evaluation samples used in this study may not fully encompass all S-CAFs present in the target projects. This limitation stems from CAFD’s bottom-up analysis strategy. For instance, in a call chain where function A calls B and B calls C, if the LLM determines that B is not a S-CAF, the analysis does not proceed to A, even if A itself could potentially be one.
Specifically, for those S-CAFs missed by the LLM during its initial pass, we employed an automated, lightweight bottom-up search along the call graph.
This search followed the same bottom-up analysis strategy defined in \$~\ref{subsec:approach}, except that it omitted the side-effect classification step.
Its sole purpose was to traverse upward from each false-negative S-CAF to identify all ancestor caller functions that serve as CAFs.
We then manually examine these upstream callers to determine whether they qualify as S-CAFs. 
We restricted this manual review to the upstream callers of false-negative S-CAFs because, under the bottom-up paradigm, any missed S-CAF must originate from a break in the analysis chain at a false-negative node.
In our analysis, we focused on the false-negative S-CAFs produced by phi and qwen, as these two models demonstrated the strongest overall performance among those evaluated. 
Notably, among the 29 false-negative S-CAFs missed by these models, their upstream callers yielded only two additional S-CAFs. 
This finding suggests that CAFD’s relatively simple bottom-up strategy is nonetheless effective in capturing the vast majority of S-CAFs in practice.

\begin{table}[htbp]
\caption{Accuracy of LLM decision.}\vspace{-2mm}
\centering
\resizebox{0.45\textwidth}{!}{%
\begin{tabular}{c|c|c|c|c|c|c|c}
\toprule
\textbf{model} & total & TP & FP & TN & FN & Prec(\%) & Recall(\%) \\
\hline
qwen & 437 & 86 & 0 & 322 & 29 & 100 & 74.8 \\
\hline
phi & 438 & 91 & 0 & 320 & 27 & 100 & 77.1  \\
\hline
llama & 127 & 32 & 0 & 61 & 34 & 100 & 48.5 \\
\hline
ds & 420 & 81 & 14 & 295 & 30 & 85.3 & 73  \\
\hline
kimi-k2 & 41 & 4 & 0 & 34 & 3 & 100 & 57.1  \\
\bottomrule
\end{tabular}
}
\vspace{-3mm}
\label{tab:afd_acc}
\end{table}

\subsubsection{Differences across epochs}~\label{subsec:difference}

Table~\ref{tab:diff_res} summarizes the results across the three epochs.
The meaning of each entry in the form ($\text{num}_1$/$\text{num}_2$) is consistent with that in Table~\ref{tab:afd_res}.
For qwen and phi, the outcomes remain identical across all epochs.
In contrast, ds and llama exhibit minor variations across several projects.
Specifically, in ds, a few allocators containing side effects are still identified in epochs 2 and 3, which can be attributed to its insufficient path understanding.
As shown in Figure~\ref{fig:af_example}(b), line 6 lies on both normal and error-handling paths, but ds incorrectly classifies it as error-handling only, resulting in a false C-CAF.
For llama, limited code understanding occasionally leads to missing more allocators.
Overall, the differences across epochs are marginal.
This is mainly because the task assigned to the LLMs is relatively simple, requiring them to determine whether the analyzed statements belong to error-handling regions.
Such regions in allocators are generally easy to identify, and most results can be inferred through propagation analysis.

\begin{table}[htbp]
\caption{Difference between epochs.}\vspace{-2mm}
\centering
\resizebox{0.45\textwidth}{!}{%
\begin{tabular}{c|c|c|c|c|c}
\toprule
\textbf{model} & \multicolumn{3}{c|}{ds} & \multicolumn{2}{c}{llama} \\
\hline
project & openssl & systemd & tmux & curl & screen \\
\hline
epoch1 & 150/34 & 21/9 & 97/7 & 15/0 & 3/3 \\
\hline
epoch2 & 145/28 & 20/8 & 96/6 & 16/1 & 2/2 \\
\hline
epoch3 & 148/31 & 20/8 & 96/6 & 16/1 & 3/3 \\
\bottomrule
\end{tabular}
}
\vspace{-3mm}
\label{tab:diff_res}
\end{table}

\subsubsection{Comparison with baseline approaches}~\label{subsec:AFD_baseline}

Since phi is the best-performing model according to \$~\ref{subsec:overall_effectiveness} and \$~\ref{subsec:acc_afd}, we use CAFD(phi) to compare against the two baseline approaches, KMeld~\cite{KMeld} and Raisin~\cite{Raisin}. 
For the outputs of KMeld and Raisin, we performed manual validation totaling 40 person-hours.
Each reported function was classified into three categories: (1) S-CAF — allocates memory
without side effects; (2) C-CAF — allocates memory but involves side effects (e.g., writing
pointer-typed values into fields of the returned buffer); and (3) non-allocator. CAFD
achieves \textbf{100\% precision} on S-CAFs. 
For KMeld and Raisin, we manually labeled each result
as S-CAF or C-CAF and report precision as
$\text{precision} = \frac{\text{S-CAF}}{\text{All}}$, where ``All'' denotes the total number
of functions detected by each tool.

\begin{figure}[t]
  \centering  \includegraphics[width=\textwidth]{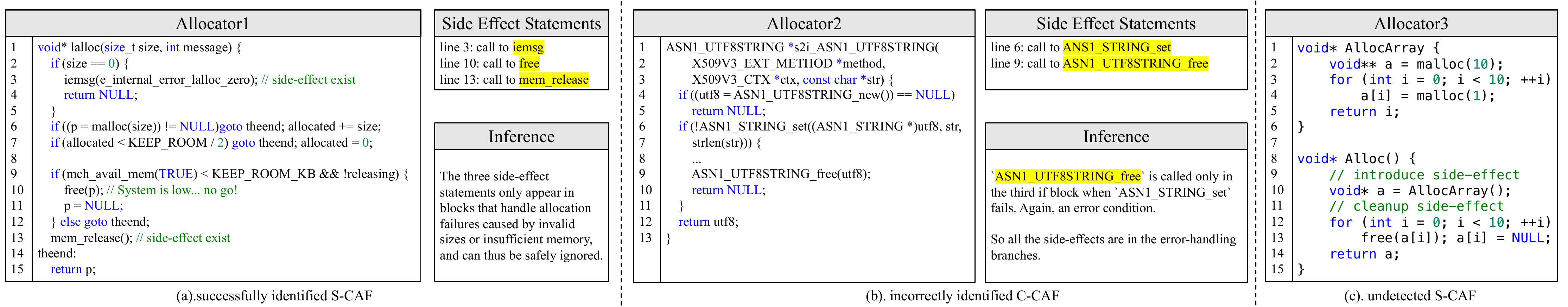}
  \vspace{-6mm}
  \caption{Allocator Examples.}
  \vspace{-4mm}
  \label{fig:af_example}
\end{figure}

Table~\ref{tab:afd_compare} summarizes the comparison between CAFD, KMeld~\cite{KMeld}, and Raisin~\cite{Raisin}. Both KMeld and Raisin show substantially lower recall (61 and 174 vs.\ 805) and lower precision than CAFD. The root cause is their aggressive allocator identification that ignores side effects, producing many C-CAFs. 
Besides, KMeld’s matching rules are coarse, excluding many true S-CAFs and misclassifying common pointer-manipulation routines (e.g., data lookup functions). Raisin further suffers from (1) dependence on manual API annotations (e.g., using \texttt{OPENSSL\_malloc} as a seed improves OpenSSL results but requires human effort), and (2) an unreliable assumption that allocators exhibit consistent contextual call-sequences (initialization/use/deallocation). Naming conventions and contexts vary widely across codebases, such heuristics yield inconsistent matches.
Nevertheless, CAFD misses approximately 31 S-CAFs, some of which are identified by KMeld and Raisin. 
For example, as shown in Figure~\ref{fig:af_example}(c), the \texttt{Alloc} function allocates 
an object that temporarily holds pointers to other objects and frees them within the same 
function, forming a side-effect creation and cleanup pattern. CAFD’s conservative strategy 
intentionally overlooks such cases.
While KMeld and Raisin are able to capture them, their more aggressive matching strategy simultaneously introduces a large number of false positives, and the resulting soundness issues make their outputs unsuitable for directly enhancing pointer analysis.

\begin{table}[htbp]
\caption{Comparison with baselines.}\vspace{-2mm}
\centering
\renewcommand{\arraystretch}{1.1}
\resizebox{\textwidth}{!}{%
\begin{tabular}{c|c|ccccccccccccccccc|c}
\toprule
\multicolumn{2}{c|}{\textbf{Project}} & bash & curl & git & h2o & htop & lighttpd & nanomq & nasm & openssl & perl & screen & systemd & teeworlds & tmux & vim & wine & P1 & \textbf{Total} \\
\midrule
\multirow{4}{*}{\cellcolor{white}{} KMeld} & All & 50 & 14 & 26 & 36 & 3 & 12 & 97 & 11 & 894 & 448 & 1 & 112 & 0 & 19 & 5 & 0 & 254  & 1946 \\
\cline{2-20}
 & \graycell S-CAF & \graycell 11 & \graycell 4 & \graycell 4 & \graycell 0 & \graycell 0 & \graycell 0 & \graycell 8 & \graycell 4 & \graycell 13 & \graycell 5 & \graycell 0 & \graycell 5 & \graycell 0 & \graycell 0 & \graycell 0 & \graycell 0 & \graycell 7 & \graycell 61 \\
\cline{2-20}
& \lightgraycell precision(\%) & \lightgraycell 22 & \lightgraycell 28.6 & \lightgraycell 15.4 & \lightgraycell 0 & \lightgraycell 0 & \lightgraycell 0 & \lightgraycell 8.2 & \lightgraycell 36.4 & \lightgraycell 1.5 & \lightgraycell 1.1 & \lightgraycell 0 & \lightgraycell 4.5 & \lightgraycell Nan & \lightgraycell 0 & \lightgraycell 0 & \lightgraycell Nan & \lightgraycell 2.8 & \lightgraycell 3.1 \\
\cline{2-20} 
& C-CAF & 14 & 3 & 7 & 5 & 2 & 4 & 18 & 2 & 368 & 16 & 0 & 4 & 0 & 7 & 1 & 0 & 156 & 607 \\
\hline

\multirow{4}{*}{Raisin} & All & 249 & 142 & 0 & 111 & 245 & 0 & 178 & 0 & 201 & 107 & 36 & 93 & 0 & 106 & 135 & 75 & 124 & 1807 \\
\cline{2-20}
& \graycell S-CAF & \graycell 16 & \graycell 11 & \graycell 0 & \graycell 1 & \graycell 9 & \graycell 0 & \graycell 16 & \graycell 0 & \graycell 24 & \graycell 6 & \graycell 4 & \graycell 3 & \graycell 0 & \graycell 23 & \graycell 36 & \graycell 7 & \graycell 18 & \graycell 174 \\
\cline{2-20}
& \lightgraycell precision(\%) & \lightgraycell 6.4 & \lightgraycell 7.7 & \lightgraycell Nan & \lightgraycell 0.9 & \lightgraycell 3.7 & \lightgraycell Nan & \lightgraycell 9.0 & \lightgraycell Nan & \lightgraycell 11.9 & \lightgraycell 5.6 & \lightgraycell 11.1 & \lightgraycell 3.2 & \lightgraycell Nan & \lightgraycell 21.7 & \lightgraycell 26.7 & \lightgraycell 9.3 & \lightgraycell 14.5 & \lightgraycell 9.6 \\
\cline{2-20}
& C-CAF & 12 & 5 & 0 & 10 & 1 & 0 & 9 & 0 & 75 & 12 & 0 & 4 & 0 & 21 & 16 & 32 & 2 & 199 \\
\hline

\grayrow \cellcolor{white} CAFD & S-CAF & \textbf{64} & \textbf{16} & \textbf{55} & \textbf{17} & \textbf{10} & \textbf{22} & \textbf{56} & \textbf{23} & \textbf{138} & \textbf{10} & \textbf{4} & \textbf{28} & \textbf{4} & \textbf{96} & \textbf{64} & \textbf{127} & \textbf{77} & \textbf{805} \\
\hline
\end{tabular}
}
\vspace{-3mm}
\label{tab:afd_compare}
\end{table}

\subsubsection{Token and time cost}~\label{subsec:afd_cost}

We present the time and cost results of CAFD on open-source programs in Table~\ref{tab:afd_cost} and on Ant Group P1 in Table~\ref{tab:afd_antgroup_cost}. 
The columns in these tables represent: \textbf{TT} – total analysis time,
\textbf{LT} – time spent on querying the LLM,
\textbf{QN} – total number of queries,
\textbf{AT/Q} – average time per query,
\textbf{IT} – total input tokens, and
\textbf{OT} – total output tokens.

Overall, the non-LLM components of CAFD incur minimal time overhead. 
For instance, even on the largest project (openssl) with qwen, the heuristic-based analysis accounts for only around 420 seconds. 
This efficiency stems from the design of CAFD, which avoids analyzing complex memory operations such as store and load.
In contrast, the LLM query time varies notably across models.
phi incurs the highest overhead, whereas llama is lightweight—detecting valid S-CAFs in bash within 7 seconds and in systemd within 42 seconds, with overhead comparable to that of small-scale models.
Unfortunately, llama’s effectiveness lags behind the other three models in terms of S-CAF identification.
Nonetheless, given the rapid development of open-source LLMs, we are optimistic that upcoming models will combine llama’s low cost with phi or qwen’s reasoning power.


\begin{table}[htbp]
\caption{Token and time cost of CAFD on open-source programs.}\vspace{-2mm}
\centering
\renewcommand{\arraystretch}{1.1}
\resizebox{\textwidth}{!}{%
\begin{tabular}{c|cccc|cccc|cccc|cccc}
\toprule
\textbf{Project} & \multicolumn{4}{c|}{bash} & \multicolumn{4}{c|}{curl} & \multicolumn{4}{c|}{git} & \multicolumn{4}{c}{h2o} \\
\hline
\textbf{model} & qwen & llama & phi & deepseek & qwen & llama & phi & deepseek & qwen & llama & phi & deepseek & qwen & llama & phi & deepseek \\
\hline
\textbf{TT(s)} & 4135 & 7 & 6946 & 622 & 742 & 2 & 1617 & 177 & 5421 & 6 & 9288 & 843 & 126 & 2 & 142 & 128 \\
\hline
\textbf{LT(s)} & 4056 & 0.1 & 6871 & 550 & 738 & 0 & 1595 & 174 & 5305 & 0.1 & 9179 & 742 & 124 & 0 & 140 & 126 \\
\hline
\textbf{QN} & 162 & 155 & 155 & 160 & 50 & 45 & 45 & 45 & 226 & 135 & 225 & 225 & 25 & 25 & 25 & 25 \\
\hline
\textbf{AT/Q(s)} & 25.0 & 0 & 37.9 & 3.4 & 14.8 & 0 & 35.4 & 3.9 & 23.5 & 0 & 40.8 & 3.3 & 5.0 & 0 & 5.6 & 5.0 \\
\hline
\textbf{IT(k)} & 85.4 & 85.1 & 90.2 & 85.4 & 21.1 & 20.8 & 22.1 & 19.5 & 90.0 & 57.7 & 103.2 & 90.4 & 7.8 & 7.5 & 9.3 & 7.8 \\
\hline
\textbf{OT(k)} & 84.9 & 0.31 & 252.5 & 93.0 & 16.3 & 0.09 & 58.3 & 20.9 & 118.0 & 0.27 & 341.9 & 111 & 12.1 & 0.24 & 20.8 & 12.0 \\
\bottomrule

\textbf{Project} & \multicolumn{4}{c|}{htop} & \multicolumn{4}{c|}{lighttpd} & \multicolumn{4}{c|}{nanomq} & \multicolumn{4}{c}{nasm} \\
\hline
\textbf{model} & qwen & llama & phi & deepseek & qwen & llama & phi & deepseek & qwen & llama & phi & deepseek & qwen & llama & phi & deepseek \\
\hline
\textbf{TT(s)} & 114 & 0 & 392 & 34 & 262 & 0 & 550 & 97 & 2700 & 2 & 2169 & 240 & 472 & 0 & 480 & 178 \\
\hline
\textbf{LT(s)} & 108 & 0 & 384 & 33 & 262 & 0 & 541 & 95 & 2626 & 0 & 2136 & 233 & 465 & 0 & 472 & 177 \\
\hline
\textbf{QN} & 10 & 5 & 10 & 10 & 20 & 20 & 20 & 20 & 146 & 50 & 70 & 60 & 16 & 15 & 15 & 15 \\
\hline
\textbf{AT/Q(s)} & 10.8 & 0 & 38.4 & 3.3 & 13.1 & 0 & 27 & 4.8 & 18.0 & 0 & 30.5 & 3.9 & 29.1 & 0 & 31.5 & 11.8 \\
\hline
\textbf{OT(k)} & 3.2 & 1.5 & 3.8 & 3.2 & 7.8 & 7.8 & 8.4 & 7.2 & 57.9 & 21.3 & 34.8 & 25.1 & 11.3 & 11.6 & 12 & 11.2 \\
\hline
\textbf{TT(k)} & 3.0 & 0.01 & 14.6 & 3.2 & 6.7 & 0.04 & 20.6 & 8.6 & 69.9 & 0.1 & 81.7 & 26.0 & 10.3 & 0.03 & 18 & 13.5 \\
\bottomrule

\textbf{Project} & \multicolumn{4}{c|}{openssl} & \multicolumn{4}{c|}{perl} & \multicolumn{4}{c|}{screen} & \multicolumn{4}{c}{systemd} \\
\hline
\textbf{model} & qwen & llama & phi & deepseek & qwen & llama & phi & deepseek & qwen & llama & phi & deepseek & qwen & llama & phi & deepseek \\
\hline
\textbf{TT(s)} & 18400 & 341 & 28485 & 3241 & 847 & 1 & 1742 & 147 & 358 & 1 & 1120 & 92 & 2498 & 42 & 6830 & 801 \\
\hline
\textbf{LT(s)} & 17977 & 0 & 27985 & 2857 & 823 & 0 & 1722 & 144 & 343 & 0 & 1107 & 90 & 2452 & 0 & 6779 & 753 \\
\hline
\textbf{QN} & 871 & 5 & 855 & 855 & 45 & 25 & 45 & 45 & 25 & 25 & 25 & 25 & 110 & 110 & 110 & 110 \\
\hline
\textbf{AT/Q(s)} & 20.6 & 0 & 32.7 & 3.2 & 18.3 & 0 & 0 & 3.2 & 13.7 & 0 & 44.3 & 3.6 & 22.3 & 0 & 61.6 & 6.9 \\
\hline
\textbf{IT(k)} & 281.0 & 2.3 & 321.7 & 284.3 & 18.9 & 9.7 & 21.5 & 18.9 & 10.2 & 11.0 & 11.7 & 10.2 & 60.2 & 63.1 & 65.3 & 60.3 \\
\hline
\textbf{OT(k)} & 446.8 & 0 & 2053.7 & 418.1 & 19.4 & 0.05 & 62.9 & 18.2 & 9.5 & 0.05 & 42.1 & 10.5 & 66.4 & 0.24 & 251.7 & 84.0 \\
\bottomrule

\textbf{Project} & \multicolumn{4}{c|}{teeworlds} & \multicolumn{4}{c|}{tmux} & \multicolumn{4}{c|}{vim} & \multicolumn{4}{c}{wine} \\
\hline
\textbf{model} & qwen & llama & phi & deepseek & qwen & llama & phi & deepseek & qwen & llama & phi & deepseek & qwen & llama & phi & deepseek \\
\hline
\textbf{TT(s)} & 100 & 0 & 133 & 11 & 3459 & 86 & 6734 & 592 & 6511 & 0 & 12774 & 1183 & 510 & 1 & 732 & 91 \\
\hline
\textbf{LT(s)} & 97 & 0 & 131 & 11 & 3374 & 1 & 6649 & 572 & 6392 & 0 & 12651 & 1071 & 491 & 0 & 711 & 87 \\
\hline
\textbf{QN} & 5 & 5 & 5 & 5 & 184 & 178 & 180 & 180 & 255 & 10 & 255 & 255 & 35 & 35 & 35 & 35 \\
\hline
\textbf{AT/Q(s)} & 19.4 & 0 & 26.2 & 2.2 & 18.3 & 0 & 36.9 & 3.2 & 25.1 & 0 & 49.6 & 4.2 & 14.0 & 0 & 20.3 & 2.5 \\
\hline
\textbf{IT(k)} & 1.4 & 1.6 & 1.7 & 1.4 & 73.8 & 74.5 & 84.1 & 73.5 & 132.7 & 4.1 & 147.7 & 133.6 & 21.1 & 20.6 & 21.3 & 21.1 \\
\hline
\textbf{OT(k)} & 1.9 & 0.01 & 4.5 & 1.5 & 85.7 & 0.02 & 244.4 & 87.2 & 138.0 & 0.02 & 475.7 & 156.6 & 13.1 & 0.07 & 15.4 & 10.8 \\
\bottomrule
\end{tabular}
}
\label{tab:afd_cost}
\end{table}

\begin{table}[htbp]
\caption{Token and time cost of CAFD on Antgroup P1.}\vspace{-2mm}
\centering
\renewcommand{\arraystretch}{1.1}
\resizebox{0.4\textwidth}{!}{%
\begin{tabular}{c|c|c|c|c|c|c}
\toprule
\textbf{model} & \textbf{TT(s)} & \textbf{LT(s)} & \textbf{QN} & \textbf{AT/Q(s)} & \textbf{IT(k)} & \textbf{OT(k)} \\
\hline
kimi-k2 & 475 & 296 & 260 & 1.1 & 131.2 & 47 \\ 
\bottomrule
\end{tabular}
}
\label{tab:afd_antgroup_cost}
\end{table}

\vspace{2mm}

\noindent
\begin{tcolorbox}[size=title, opacityfill=0.1, nobeforeafter]
\textbf{ANSWER:} \textit{CAFD, when combined with strong LLMs, can effectively identify S-CAFs with low overhead. Compared to conservative heuristic-based methods that often miss S-CAFs, LLMs help bridge the gap through semantic reasoning.}
\compactline
\end{tcolorbox}

\subsection{RQ2: Effectiveness in Enhancing Pointer Analysis}

We evaluate the effectiveness of the detected S-CAFs in enhancing pointer analysis using
the metrics defined in \$~\ref{subsec:emprical_res}. 
In particular, when computing $\text{ARR}$, the group of pointers $P$ is defined following \$~\ref{subsec:metrics} as the set of allocation receivers corresponding to all heap objects considered in the LLM-enhanced analysis. 
Formally, $P = \{\text{ar}(o) \mid \forall o \in (A \cup R)\}$, where $A$ and $R$ denote the sets of allocation and return objects obtained by combining phi with CAFD to enhance the pointer analysis.

Following \$~\ref{subsec:AFD_baseline}, we use the S-CAFs identified by the phi, the
best-performing LLM, to compare against the pure heuristic strategy~(denoted as \textbf{CAFD-H}), in which
LLM-based side-effect ignoring analysis is disabled, and a 1-callsite-sensitive baseline
(abbreviated as \textbf{1ctx}).
The 1ctx strategy performs global context-sensitive modeling across the entire program. For example, as illustrated in Figure~\ref{fig:example}, it models memory objects $(c_1, o)$ and $(c_2, o)$ in \texttt{array\_create} and \texttt{make\_bare\_word}, respectively. 
This results in $\text{pbs}_e((c_1, o)) = \{(c_1, \text{temp}), (\_, r)\}$ and $\text{as}_e((\_, r)) = \{(c_1, \text{temp})\}$, showing how context-sensitive pointers and objects help reduce redundant or spurious alias relationships.
Table~\ref{tab:avg_res} summarizes the average results of the three strategies across all
projects, including $\text{PRR}_1$, $\text{PRR}_2$, and $\text{ARR}$, as well as the average
heap object expansion ratio (denoted as \textbf{ER}). 
The \textbf{ER} score measures the growth in THOC and is computed as $\frac{\text{num}_2}{\text{num}_1}$, where $\text{num}_1 \rightarrow \text{num}_2$ indicates
the change in the number of heap objects before and after applying the corresponding strategy.

As shown in Table~\ref{tab:avg_res}, all strategies lead to improvements in
$\text{PRR}_1$, $\text{PRR}_2$, $\text{ARR}$, and \textbf{ER} scores, indicating that the
identified S-CAFs help distribute pointers across more heap objects, reduce spurious
cross-context alias relationships, and expand the tracked heap space. Compared with the
heuristic-based approach, the LLM-enhanced strategies achieve higher average ER, PRR, and
ARR scores, demonstrating that incorporating LLMs enables more advanced S-CAF identification
and thus a more effective pointer analysis enhancement.
We further compare CAFD with 1ctx. CAFD consistently yields a higher ER than 1ctx, indicating that it models more heap objects. This is particularly important in real-world projects where CAFs are frequently used and may be nested across multiple layers down to system allocation APIs, which the 1ctx
strategy alone cannot fully cover. In terms of $\text{PRR}_1$ and $\text{PRR}_2$, direct
comparison between CAFD and 1ctx is not straightforward: CAFD focuses on local improvements for custom S-CAFs, whereas 1ctx performs a global context-sensitive analysis covering the entire program. Nevertheless, considering $\text{ARR}$, CAFD outperforms 1ctx, indicating that CAFD provides greater benefits for local heap-related analyses.

\vspace{-1mm}
\begin{table}[htbp]
\caption{Average results of three strategies.}\vspace{-2mm}
\resizebox{0.35\textwidth}{!}{
\begin{tabular}{c|c|c|c|c}
\hline
\textbf{strategy} & ER & $\text{PRR}_1$(\%) & $\text{PRR}_2$(\%) & ARR(\%) \\
\hline
CAFD & 38.8 & 54.3 & 17.0 & 41.5 \\
CAFD-H & 6.5 & 27.3 & 4.7 & 16.1 \\
1ctx & 23.2 & 41.4 & Nan & 27.2 \\
\hline
\end{tabular}
}\vspace{-3mm}
\label{tab:avg_res}
\end{table}

\begin{table*}[htbp]
\caption{Overall results of pointer analysis. CAFD \&-H denotes the results common to qwen and heuristic. ``Nan'' denotes that A set is empty. $\uparrow$ indicates that higher values correspond to better results.}
\centering
\vspace{-1mm}
\resizebox{0.95\textwidth}{!}{%
\begin{tabular}{c|c|c|c|c|c|c|c|c|c}
\hline
project & Strategy & THOC & SUP & $\text{PC}_1$ & $\text{PRR}_1$(\%) $\uparrow$ & $\text{PC}_2$ & $\text{PRR}_2$(\%) $\uparrow$ & ANC & ARR(\%) $\uparrow$ \\
\hline
\multirow{2}{*}{bash} & CAFD & 49 $\rightarrow$ 1.25K & 42 & 33.48K $\rightarrow$ 11.2K & 66.5 & 9.97K $\rightarrow$ 6.27K & 37.2 & 32.69K $\rightarrow$ 11.03K & 66.3 \\
 & 1ctx & 49 $\rightarrow$ 1.31K & 0 & 13.3K $\rightarrow$ 6.6K & 50.6 & Nan & Nan & 32.69K $\rightarrow$ 15.99K & 51.1 \\
\hline

\multirow{3}{*}{curl} & CAFD & 69 $\rightarrow$ 151 & 66 & 27.93K $\rightarrow$ 14.02K & 49.8 & 23.64K $\rightarrow$ 23.29K & 1.5 & 25.38K $\rightarrow$ 20.6K & 18.9 \\
 & CAFD-H & 69 $\rightarrow$ 151 & 66 & 27.93K $\rightarrow$ 18.73K & 33.0 & 23.7K $\rightarrow$ 23.7K & 0.0 & 25.38K $\rightarrow$ 20.83K & 18.0 \\
 & 1ctx & 69 $\rightarrow$ 191 & 0 & 23.82K $\rightarrow$ 15.44K & 35.2 & Nan $\rightarrow$ Nan & 0.0 & 25.38K $\rightarrow$ 21.27K & 16.2 \\
\hline

\multirow{2}{*}{h2o} & CAFD \&-H & 106 $\rightarrow$ 366 & 93 & 16.35K $\rightarrow$ 2.57K & 84.3 & 12.29K $\rightarrow$ 9.56K & 21.9 & 15.06K $\rightarrow$ 10.58K & 29.8 \\
 & 1ctx & 106 $\rightarrow$ 491 & 0 & 12.79K $\rightarrow$ 7.91K & 39.1 & Nan & Nan & 15.06K $\rightarrow$ 11.21K & 25.6 \\
 \hline

\multirow{3}{*}{htop} & CAFD & 6 $\rightarrow$ 165 & 1 & 10.74K $\rightarrow$ 7.19K & 33.1 & 1 $\rightarrow$ 1 & 0.0 & 10.68K $\rightarrow$ 7.14K & 33.1 \\
 & CAFD-H & 6 $\rightarrow$ 145 & 2 & 10.77K $\rightarrow$ 7.4K & 31.3 & 5.24K $\rightarrow$ 5.11K & 2.6 & 10.68K $\rightarrow$ 7.77K & 27.2 \\
 & 1ctx & 6 $\rightarrow$ 146 & 0 & 10.73K $\rightarrow$ 7.5K & 30.1 & Nan & Nan & 10.68K $\rightarrow$ 8.7K & 18.2 \\
\hline

\multirow{3}{*}{git} & CAFD & 16 $\rightarrow$ 1.56K & 7 & 121.11K $\rightarrow$ 77.86K & 35.7 & 96.92K $\rightarrow$ 79.53K & 17.9 & 120.73K $\rightarrow$ 78.68K & 34.8 \\
 & CAFD-H & 16 $\rightarrow$ 61 & 14 & 111.01K $\rightarrow$ 84.75K & 23.7 & 108.14K $\rightarrow$ 108.14K & 0.0 & 120.73K $\rightarrow$ 120.26K & 0.4 \\
 & 1ctx & 16 $\rightarrow$ 1.69K & 0 & 110.53K $\rightarrow$ 62K & 43.9 & Nan & Nan & 120.73K $\rightarrow$ 92.87K & 23.1 \\
\hline

\multirow{2}{*}{lighttpd} & CAFD \&-H & 10 $\rightarrow$ 77 & 7 & 12.2K $\rightarrow$ 7.51K & 38.5 & 10.63K $\rightarrow$ 7.5K & 29.4 & 12.06K $\rightarrow$ 7.51K & 37.7 \\
 & 1ctx & 10 $\rightarrow$ 104 & 0 & 11.1K $\rightarrow$ 4.47K & 59.7 & Nan & Nan & 12.06K $\rightarrow$ 9.48K & 21.4 \\
\hline

\multirow{3}{*}{nanomq} & CAFD & 284 $\rightarrow$ 1.36K & 267 & 66.48K $\rightarrow$ 45.47K & 31.6 & 26.88K $\rightarrow$ 22.11K & 17.7 & 56.85K $\rightarrow$ 37.29K & 34.4 \\
 & CAFD-H & 284 $\rightarrow$ 457 & 273 & 46.15K $\rightarrow$ 33.38K & 27.7 & 28.2K $\rightarrow$ 27.71K & 1.7 & 56.85K $\rightarrow$ 51.67K & 9.1 \\
 & 1ctx & 284 $\rightarrow$ 960 & 0 & 29.25K $\rightarrow$ 19.45K & 35.5 & Nan & Nan & 56.85K $\rightarrow$ 46.45K & 18.3 \\
\hline

\multirow{3}{*}{nasm} & CAFD & 5 $\rightarrow$ 245 & 3 & 14.59K $\rightarrow$ 3.98K & 72.8 & 13.66K $\rightarrow$ 10.85K & 20.7 & 14.58K $\rightarrow$ 4.06K & 72.2 \\
& CAFD-H & 5 $\rightarrow$ 245 & 3 & 14.59K $\rightarrow$ 4.01K & 72.5 & 13.66K $\rightarrow$ 10.85K & 20.6 & 14.58K $\rightarrow$ 4.09K & 72.0 \\
& 1ctx & 5 $\rightarrow$ 169 & 0 & 14.03K $\rightarrow$ 5.44K & 61.2 & Nan & Nan & 14.58K $\rightarrow$ 6.66K & 54.3 \\
\hline

\multirow{2}{*}{openssl} & CAFD & 8 $\rightarrow$ 1.95K & 6 & 171.3K $\rightarrow$ 84.79K & 50.5 & 94.67K $\rightarrow$ 90.92K & 4.0 & 170.9K $\rightarrow$ 85.8K & 49.8 \\
 & 1ctx & 8 $\rightarrow$ 484 & 0 & 117.36K $\rightarrow$ 89.19K & 24.0 & Nan & Nan & 170.9K $\rightarrow$ 134.5K & 21.3 \\
\hline

\multirow{2}{*}{perl} & CAFD & 47 $\rightarrow$ 271 & 40 & 92.22K $\rightarrow$ 60.99K & 33.9 & 83.23K $\rightarrow$ 81.97K & 1.5 & 91.38K $\rightarrow$ 68.79K & 24.7 \\
 & 1ctx & 47 $\rightarrow$ 381 & 0 & 84.6K $\rightarrow$ 50.76K & 40.0 & Nan & Nan & 91.38K $\rightarrow$ 68.35K & 25.2 \\
\hline

\multirow{2}{*}{screen} & CAFD \&-H & 84 $\rightarrow$ 145 & 80 & 27.63K $\rightarrow$ 11.76K & 57.4 & 20.78K $\rightarrow$ 20.78K & 0.0 & 23.85K $\rightarrow$ 16.74K & 29.8 \\
 & 1ctx & 84 $\rightarrow$ 443 & 0 & 21.11K $\rightarrow$ 8.53K & 59.6 & Nan & Nan & 23.85K $\rightarrow$ 17.86K & 25.1 \\
\hline

\multirow{3}{*}{systemd} & CAFD & 3.85K $\rightarrow$ 5.58K & 3.69K & 228.35K $\rightarrow$ 49.08K & 78.5 & 192.59K $\rightarrow$ 188.43K & 2.2 & 217.46K $\rightarrow$ 175.11K & 19.5 \\
 & CAFD-H & 3.85K $\rightarrow$ 5.35K & 3.79K & 229.73K $\rightarrow$ 108.95K & 52.6 & 188.62K $\rightarrow$ 184.41K & 2.2 & 217.46K $\rightarrow$ 186.54K & 14.2 \\
 & 1ctx & 3.85K $\rightarrow$ 15.75K & 0 & 194.08K $\rightarrow$ 120.13K & 39.1 & Nan & Nan & 217.46K $\rightarrow$ 166.14K & 23.6 \\
\hline

\multirow{2}{*}{teeworlds} & CAFD \&-H & 82 $\rightarrow$ 156 & 76 & 50.17K $\rightarrow$ 18.67K & 62.8 & 32.29K $\rightarrow$ 32.29K & 0.0 & 40.35K $\rightarrow$ 26.52K & 34.3 \\
 & 1ctx & 82 $\rightarrow$ 244 & 0 & 35.59K $\rightarrow$ 20.08K & 43.6 & Nan & Nan & 40.35K $\rightarrow$ 30.02K & 25.6 \\
\hline

\multirow{2}{*}{tmux} & CAFD & 18 $\rightarrow$ 630 & 14 & 36.08K $\rightarrow$ 16.87K & 53.2 & 23.39K $\rightarrow$ 20.07K & 14.2 & 30.06K $\rightarrow$ 16.87K & 43.9 \\
 & 1ctx & 18 $\rightarrow$ 617 & 0 & 26.21K $\rightarrow$ 17.64K & 32.7 & Nan & Nan & 30.06K $\rightarrow$ 21.16K & 29.6 \\
\hline

\multirow{2}{*}{vim} & CAFD & 30 $\rightarrow$ 1.78K & 27 & 184.34K $\rightarrow$ 112.98K & 38.7 & 53.07K $\rightarrow$ 40.77K & 23.2 & 182.49K $\rightarrow$ 111.89K & 38.7 \\
 & 1ctx & 30 $\rightarrow$ 173 & 0 & 66.19K $\rightarrow$ 44.94K & 32.1 & Nan & Nan & 182.49K $\rightarrow$ 146.72K & 19.6 \\
\hline

\multirow{3}{*}{wine} & CAFD & 94 $\rightarrow$ 989 & 66 & 59.52K $\rightarrow$ 36.29K & 39 & 37.74K $\rightarrow$ 37.03K & 1.9 & 56.47K $\rightarrow$ 34.54K & 38.8 \\
 & CAFD-H & 94 $\rightarrow$ 777 & 66 & 59.52K $\rightarrow$ 37.3K & 39 & 37.42K $\rightarrow$ 37.03K & 1.9 & 56.47K $\rightarrow$ 39.2K & 30.6 \\
 & 1ctx & 94 $\rightarrow$ 603 & 0 & 44.23K $\rightarrow$ 28.04K & 36.6 & Nan & Nan & 56.47K $\rightarrow$ 46.76K & 17.2 \\
\hline

\multirow{2}{*}{P1} & CAFD & 17 $\rightarrow$ 1.42K & 15 & 46.77K $\rightarrow$ 1.14K & 97.6 & 2.14K $\rightarrow$ 108 & 95 & 46.3K $\rightarrow$ 1.22K & 97.6 \\
 & 1ctx & 17 $\rightarrow$ 932 & 0 & 7.39K $\rightarrow$ 4.2K & 43.2 & Nan & Nan & 46.3K $\rightarrow$ 23.66K & 48.9 \\
\hline
\end{tabular}%
}
\label{tab:pts_res}
\end{table*}

Table~\ref{tab:pts_res} presents the results of the three strategies across 17 projects. Focusing on the heuristic strategy~(CAFD-H), we observe that it fails to identify any S-CAFs in
\texttt{screen} and \texttt{tmux}, resulting in no improvement. 
In \texttt{bash}, \texttt{openssl}, and \texttt{perl}, although CAFD-H identifies a few S-CAFs, these are not invoked and thus do not contribute to pointer analysis enhancement. 
In contrast, the LLM-enhanced strategy detects multiple effective S-CAFs that significantly improve pointer analysis, demonstrating the effectiveness of CAFD when combined with LLM-based enhancements.
Comparing CAFD with 1ctx, we find that, for all projects except \texttt{systemd}, CAFD achieves higher $\text{ARR}$ scores, indicating that CAFD is more efficient for heap-related analyses. 
Although 1ctx performs better on \texttt{systemd}, Table~\ref{tab:cost_comparison} shows that its computational overhead is significantly higher 
than CAFD. 
From the perspective of THOC, CAFD models more memory objects than 1ctx in \texttt{htop}, \texttt{nanomq}, \texttt{nasm}, \texttt{openssl}, \texttt{tmux}, \texttt{vim}, \texttt{wine}, and \texttt{P1}, whereas 1ctx models more objects in the remaining nine projects. 
This difference primarily arises from the distinct design goals: CAFD performs local enhancements focused on S-CAFs, while 1ctx applies global context-sensitive precision improvements. 
Nevertheless, these two approaches are complementary and can be seamlessly combined to achieve even higher-precision pointer analysis.

\noindent\textbf{Time And Memory Cost}.
Table~\ref{tab:cost_comparison} presents a comparison of the consumed time~(in seconds) 
and memory~(in MB) for the three strategies: pure context-insensitive analysis (base), CAFD, and 1ctx. 
Enhancing pointer analysis with CAFD leads to a moderate increase in runtime, with the total analysis time rising by approximately 1.4×, which is acceptable. In terms of memory cost, CAFD incurs roughly 1.3× the memory usage of the base analysis. 
In contrast, 1ctx introduces a substantial overhead, with runtime increasing by about 26× 
and memory usage growing more than 2.5× compared to the base analysis. Examining individual  projects, we find that the largest time overhead occurs in large projects such as 
\texttt{systemd}, \texttt{openssl}, \texttt{git}, \texttt{perl}, and \texttt{vim}, 
indicating that traditional context-sensitive strategies suffer more severe performance degradation on large-scale programs. 
The relatively low overhead of CAFD stems from its design: it focuses solely on enhancing precision by identifying additional allocation sources through S-CAFs, without introducing extra context elements, thereby achieving better scalability.

\vspace{2mm}

\noindent
\begin{tcolorbox}[size=title, opacityfill=0.1, nobeforeafter]
\textbf{ANSWER:} \textit{CAFD-generated S-CAFs enhance pointer analysis precision with minimal overhead. By increasing the number of modeled heap objects across contexts, they reduce aliasing and points-to imprecision. The LLM-enhanced CAFD identifies more effective S-CAFs than the heuristic-based approach, leading to greater overall improvement.}
\compactline
\end{tcolorbox}

\begin{table}[htbp]
\centering
\caption{Comparison of time~(s) and memory cost~(MB).}\vspace{-3mm}
\resizebox{\textwidth}{!}{
\begin{tabular}{c|ccccccccccccccccc|c}
\hline
\textbf{Project} & bash & curl & h2o & htop & git & lighttpd & nanomq & nasm & openssl & perl & screen & systemd & teeworlds & tmux & vim & wine & P1 & \textbf{total} \\
\hline
CI time & 19.6 & 79.7 & 38.1 & 3.3 & 1.04K & 1.6 & 112.3 & 33.3 & 1.42K & 219.6 & 13.5 & 1.30K & 82.2 & 45.2 & 547.3 & 43.2 & 37.1 & 5.05K \\
CAFD time & 43.6 & 73.6 & 38.1 & 6.5 & 1.55K & 2.5 & 196.0 & 26.0 & 1.76K & 289.5 & 15.2 & 1.31K & 80.4 & 81.3 & 1.55K & 49.7 & 17.6 & 7.09K  \\
1ctx time & 89.7 & 125.1 & 64.2 & 23.1 & 27.38K & 3.2 & 431.5 & 66.1 & 11.06K & 5.48K & 29.2 & 79.48K & 161.6 & 167.4 & 6.18K & 611.7 & 68.2 & 131.77K \\
\hline

CI mem & 575.9 & 1.14K & 656.6 & 175.8 & 19.94K & 105.5 & 2.61K & 2.08K & 25.59K & 2.50K & 454.0 & 20.78K & 1.87K & 899.4 & 10.88K & 1.32K & 680.2 & 92.25K  \\
CAFD mem & 1.07K & 1.19K & 673.9 & 235.1 & 36.83K & 131.5 & 3.42K & 1.58K & 24.04K & 2.97K & 487.6 & 23.56K & 2.02K & 1.41K & 18.85K & 1.71K & 696.1 & 120.87K \\
1ctx time & 2.11K & 1.87K & 1.18K & 367.5 & 44.67K & 254.8 & 5.76K & 2.45K & 29.25K & 17.85K & 642.4 & 95.03K & 789.6 & 1.90K & 23.37K & 3.87K & 1.26K & 232.61K \\
\hline
\end{tabular}
}
\label{tab:cost_comparison}
\end{table}

\subsection{RQ3: Effectiveness in Downstream Tasks}

We apply the enhanced pointer analysis to two downstream tasks: indirect-call analysis~(\S~\ref{subsec:icall}, simply denoted as icall) and value-flow-based bug detection~(\S~\ref{subsec:bug_detect}). 
For indirect calls, the goal is to resolve the target set $\text{pts}(f)$ for a given call $f(v_1, ..., v_n)$, directly relying on the precision of the function pointer.
For bug detection, pointer analysis results are used to build memory SSA~\cite{memSSA} and a value-flow graph~\cite{VFG}, which are traversed by checkers with predefined rules.
We evaluate CAFD on two detectors, Saber~\cite{Saber} and Pinpoint~\cite{pinpoint}, which together cover four bug types and five checkers: both tools detect memory leaks, Saber additionally checks double free, and Pinpoint checks null dereference and use-after-free.

\subsubsection{Indirect Call Analysis}~\label{subsec:icall}

\begin{figure}[t]
  \centering  \includegraphics[width=0.8\textwidth]{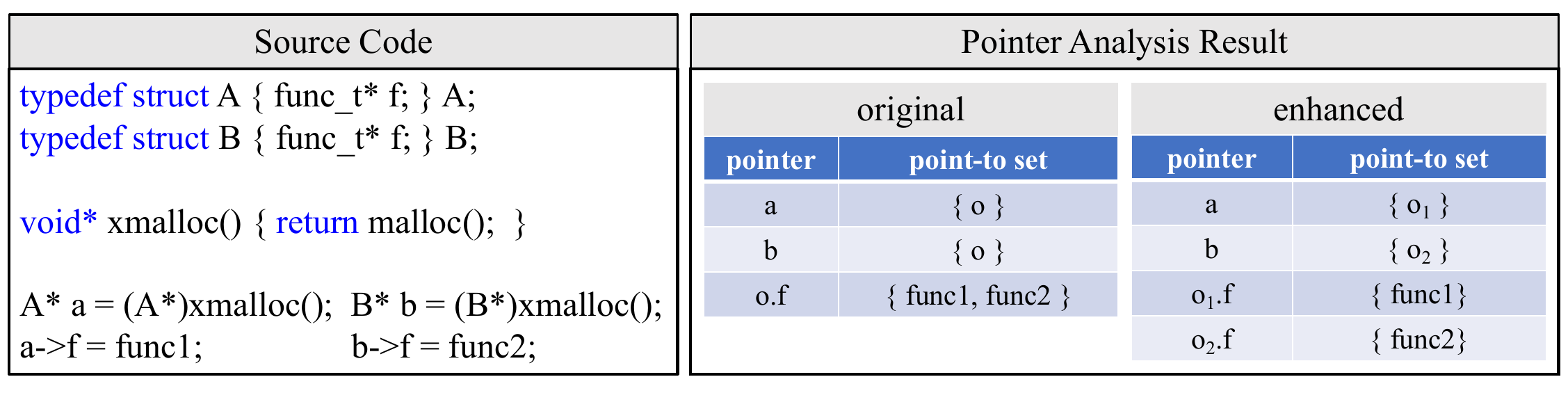}
  \vspace{-4mm}
  \caption{Example of refining target set.}
  \label{fig:icall}
  \vspace{-4mm}
\end{figure}

Across all 17 projects, SVF analyzes a total of 2,039 icalls. 
Among them, pointer analysis enhanced by CAFD improves the resolution of 284 icalls across 6 projects, reducing their average target-set size from 142.3 to 65.6.
Table~\ref{tab:icall_res} summarizes the statistics. 
Specifically, Figure~\ref{fig:icall} illustrates the refinement mechanism. Without enhancement, only one object $o$ is modeled, so both \texttt{a->f = func1} and \texttt{b->f = func2} store targets into $o.f$, leading to incorrect call targets. 
By modeling CAF \texttt{xmalloc}, our approach separates the heap objects pointed to by \texttt{a} and \texttt{b}, refining the target set.
While MLTA~\cite{MLTA} refines analysis via struct-field sensitivity, its effectiveness is limited by type escape, leaving many icalls unresolved. 
While CAFD-enhanced pointer analysis improves indirect call resolution, the gains remain moderate due to two factors: our lightweight CAF modeling prioritizes efficiency over handling complex allocation patterns like memory pools, and a prior study~\cite{Cocktail} shows that function pointers are predominantly propagated via global variables or function parameters, with relatively few stored in heap objects where our enhancements are most applicable.
Future work could explore more sophisticated CAF identification, though this would require balancing analysis cost against potential precision gains.

\begin{table}[h]
\centering
\caption{Results for icall analysis. \textbf{TN}: total icalls analyzed; \textbf{ON}: icalls optimized by CAFD. \textbf{OA}/\textbf{EA}: average targets per optimized icall under original/enhanced pointer analysis.}\vspace{-1mm}
\resizebox{0.5\textwidth}{!}{
\begin{tabular}{c|cccccc|c}
\hline
\textbf{Project} & git & htop & nanomq & vim & h2o & P1 & \textbf{total} \\
\hline
TN & 20 & 79  & 83 & 1035 & 259 & 563 & 2039 \\
ON & 4 & 50 & 1 & 41 & 114 & 74 & 284  \\
OA & 243.2  & 104.8  & 7 & 80 & 89.5 & 343.7 & 142.3  \\
EA & 235.2 & 41 & 6 & 21.9 & 64.5 & 99.6 & 65.6 \\
\hline
\end{tabular}
}
\label{tab:icall_res}
\end{table}

\subsubsection{Bug Detection}~\label{subsec:bug_detect}

This section investigates how CAFD-enhanced pointer analysis impacts bug detection. We integrate CAFD into two detectors: Saber~\cite{Saber}, which is built on top of SVF, and Pinpoint~\cite{pinpoint}, and evaluate their effect on detecting memory leak~(\textbf{ML}), double-free~(\textbf{DF}), use-after-free~(\textbf{UAF}), and null-pointer dereference~(\textbf{NPD}). 
To isolate the influence of enhancement, we exclude bug reports that remain unchanged and focus on the differing ones.
Table~\ref{tab:detection} summarizes the statistics of bug reports that differ before and after enhancement, including pruned and newly generated reports.
We spent approximately 80 person-hours on manual validation. Specifically, (1) for ML reports, we checked whether the allocated memory was left unfreed within the same scope without escaping; if it escaped, we searched globally for potential free sites—reports with such sites were marked as false positives. For other bug types, we examined whether a feasible path existed from the source to the sink.
(2) We further inspected the latest versions of the projects on platforms such as GitHub; if the corresponding code locations contained potential fix patterns, the report was labeled as a true positive.
As shown in Table~\ref{tab:detection}, after enhancement, all pruned reports were false positives (denoted as \textbf{OFP}, original FPs), while the newly generated reports included both true positives (\textbf{ETP}, enhanced TPs) and false positives (\textbf{EFP}).

In summary, for Saber, the enhancement eliminates 288 false positives, while introducing 6 true positives and 767 false positives.
For Pinpoint, it removes 234 false positives, introduces 23 true positives, and adds 124 false positives.
Among the 29 newly discovered bugs, 12 have already been fixed in the latest versions, 11 were confirmed and fixed after being reported to the developers, and the remaining 6 are awaiting developer response.
The enhanced pointer analysis provides the most significant improvement for ML detection: among the 29 newly discovered bugs, 26 are ML and 5 are NPD, and their distribution across the benchmark projects is detailed in Table~\ref{tab:bug_distribution}.
For DF, some false positives are removed, but more new ones are introduced, indicating that although additional allocation sources are identified, verifying DF-specific value flow path conditions remains challenging.
UAF detection remains unchanged, for reasons similar to those affecting DF detection, its checker logic is more complex than ML, limiting the impact of the enhancement.

\begin{figure}[t]
  \centering  \includegraphics[width=0.8\textwidth]{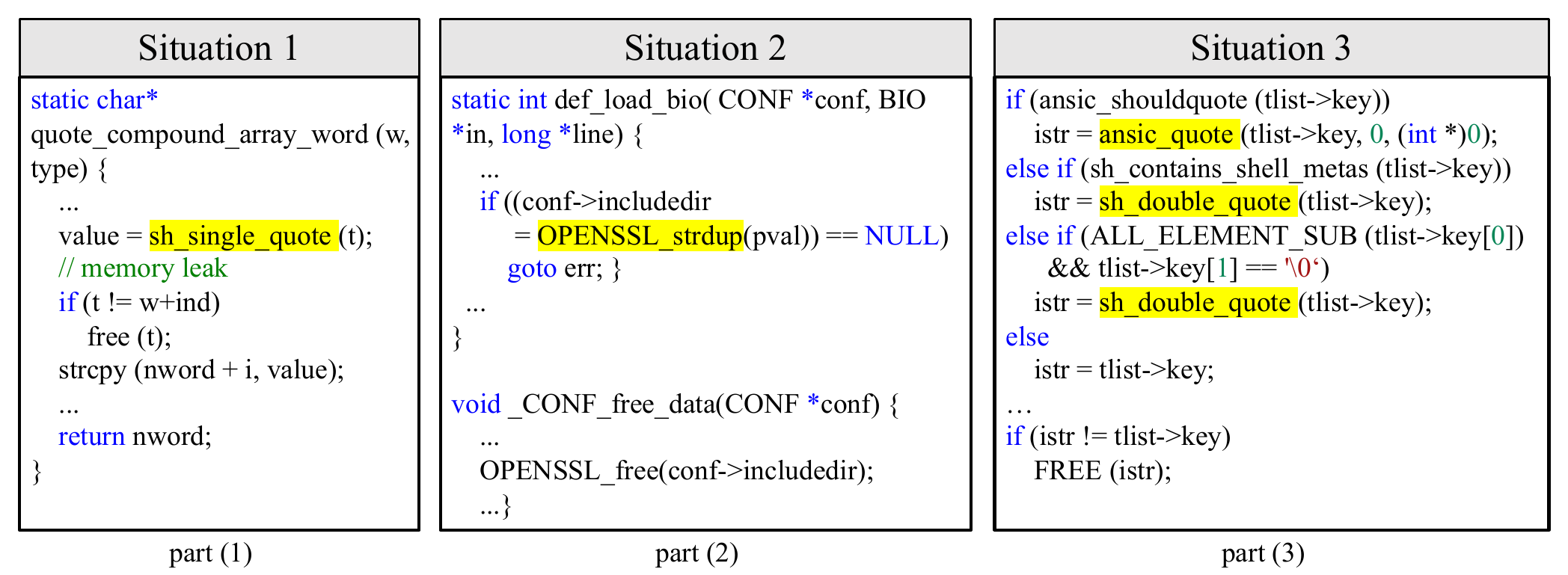}
  \vspace{-4mm}
  \caption{Bug examples for the three situations. Yellow highlights indicate the reported allocation sources.}
  \vspace{-4mm}
  \label{fig:bug_example}
\end{figure}

To better understand the effects of the enhanced pointer analysis, we conduct case studies on the three situations — \textbf{1.discovering additional true positives}, \textbf{2.eliminating false positives}, and \textbf{3.introducing new false positives} — to analyze their underlying causes.
Figure~\ref{fig:bug_example} presents three representative examples illustrating these situations.
The observed causes are generally applicable across all four bug types considered in this work.
\textbf{1.Discover new TPs:} by identifying more S-CAFs, CAFD enables Saber to model additional allocation sources. 
As shown in Figure~\ref{fig:bug_example} part (1), \texttt{sh\_single\_quote} is one such example. 
Although Saber implements a simple heuristic to detect custom allocation wrappers, it fails to recognize this function. 
For Pinpoint, this effect is even more pronounced, mainly due to the limitations of its summary-based analysis strategy.
\textbf{2.Prune FPs:} the enhanced analysis helps eliminate certain false positives because by modeling S-CAFs, it preserves value-flow paths across complex calling contexts. 
This enables recovery of previously missing flows. 
For instance, as shown in Figure~\ref{fig:bug_example}(2), the value flow from \texttt{conf->includedir} in \texttt{def\_load\_bio} to \texttt{CONF\_free\_data} is correctly recovered, thereby suppressing a spurious ML report.
\textbf{3.Introduce new FPs:} the remaining false positives mainly stem from limitations in Saber’s detection rules:
(1) Insufficient path sensitivity. As shown in Figure~\ref{fig:bug_example} part (3), \texttt{istr} may point to heap objects returned by S-CAFs like \texttt{ansic\_quote} or \texttt{sh\_double\_quote}. 
Due to the path condition \texttt{istr != tlist->key}, Saber fails to recognize that these objects can be freed in any execution path, leading to false positives. This accounts for approximately 65\% of memory leak FPs, with similar cases in double frees.
(2) Imprecise value-flow construction. Complex calling contexts can cause interprocedural value flows to be missed, further contributing to false positives.

Overall, CAFD enables Saber and Pinpoint to identify more source points, thereby uncovering additional potential memory issues. However, the false-positive challenge remains, largely due to inherent limitations in the checkers’ detection mechanisms rather than in the precision of source identification. 
A promising direction is to integrate an LLM-based agent to analyze the code snippets referenced in bug reports and automatically filter out false positives. 
To further improve detection accuracy, advances in both path-sensitive reasoning and precise source-to-sink value-flow construction are still essential.

\vspace{2mm}

\noindent
\begin{tcolorbox}[size=title, opacityfill=0.1, nobeforeafter]
\textbf{ANSWER:}\textit{ Overall, CAFD can enhance downstream tasks driven by pointer analysis. 
Future work could be focused on improving CAF identification in complex scenarios and incorporating advanced value-flow analysis to further boost downstream effectiveness.}
\compactline
\end{tcolorbox}

\begin{table}[h]
\centering
\caption{Overall results of bug detection.}
\vspace{-1mm}
\resizebox{0.38\textwidth}{!}{
\begin{tabular}{c|c|c|c|c}
\hline
Tool & Bug Type & OFP & ETP & EFP\\
\hline
\multirow{2}{*}{Saber} & ML & 258 & 6 & 697 \\
\cline{2-5}
& DF & 30 & 0 & 70 \\
\hline
\multirow{2}{*}{Pinpoint} & ML & 1 & 18 & 67 \\
\cline{2-5}
& NPD & 233 & 5 & 57 \\
\hline
\end{tabular}
}
\label{tab:detection}
\end{table}

\begin{table}[htbp]
\caption{Bug distribution across projects}
\centering
\resizebox{0.65\textwidth}{!}{%
\begin{tabular}{c|c|c|c|c|c|c|c|c|c}
\hline
project & bash & openssl & perl & \multicolumn{2}{c|}{tmux} & h2o & wine & \multicolumn{2}{c}{P1} \\
\hline
bug type & ML & ML & NPD & ML & NPD & ML & ML & ML & NPD \\
\hline
bugs & 1 & 10 & 1 & 6 & 1 & 1 & 3 & 3 & 3 \\
\hline
\end{tabular}%
}
\label{tab:bug_distribution}
\end{table}

\section{Discussion}

While CAFD achieves promising results, several aspects could benefit from further exploration.
\textbf{First}, CAFD currently employs a simple S-CAF detection strategy to remain lightweight, which may miss cases involving more complex side effects. 
For example, S-CAFs that allocate a heap object to a local variable and immediately free it. 
Although such cases are not handled in this work, they constitute only a small proportion of all S-CAFs and therefore have minimal impact on CAFD’s overall effectiveness.
\textbf{Second}, real-world projects often implement C-CAFs.
These functions may exhibit behaviors such as memory pooling or garbage-collection–style management, which are more intricate than standard heap allocation but can still be approximated as malloc-like in many cases.
In addition, objects returned by such allocators may have hierarchical structures~\cite{Goshawk} or internal pointer relationships that cannot be captured by a simple malloc abstraction.
Improving the detection and modeling of these CAFs would further enhance CAFD’s precision and scalability, and therefore represents an important direction for future work.
\textbf{Third}, Bug detection is an important downstream task of pointer analysis. Our evaluation (\S~\ref{subsec:bug_detect}) reveals a counterintuitive finding: although modeling CAFs improves pointer analysis precision and uncovers bugs that were previously undetectable, such improvements alone do not lead to a substantial boost in overall bug detection effectiveness, particularly in reducing false positives.
This suggests that future research should investigate how to better exploit refined pointer information, for example by developing more sophisticated checkers or leveraging LLM-based agents to filter false alarms and further improve detection precision.
\textbf{Finally}, CAFD currently focuses on C projects. Given the significant differences in program features and memory models, its effectiveness on languages like Java may be limited. 
Future work should aim to improve pointer analysis in other programming languages.

\section{Threat To Validity}

\textbf{Evaluation Metrics}.
A key threat stems from the absence of standardized metrics for evaluating C/C++ pointer analyses. 
Following common practice, we assess the effectiveness of our approach using points-to set sizes and alias-set sizes. 
In particular, we examine how the enhanced analysis affects (1) the average points-to set size of heap objects and (2) the average alias-set size of allocation receivers.
A fundamental challenge is that real-world programs do not provide ground-truth heap-receiver sets. To alleviate this issue, we apply a combination-set principle: for each allocation site, we approximate the possible receiver set by taking the union of receivers collected across all analysis configurations. This union acts as an upper bound and serves as a stable reference for evaluating whether our enhanced analysis moves toward or away from the broadest plausible aliasing relationship. 
While this approximation is not perfect, it provides a practical and consistent basis for comparison in the absence of oracle information. 
In future work, we plan to combine dynamic information collection with manual inspection to construct larger and more accurate receiver sets, further improving the evaluation.

\noindent \textbf{Soundness of S-CAF Identification}.
Another threat concerns the soundness of S-CAF identification, since the identification process involves LLMs and may introduce misclassifications. 
We take several steps to mitigate this risk.
First, the tasks assigned to the LLMs are deliberately simple and do not require complex reasoning, reducing the likelihood of hallucinations or unstable behavior.
Second, we employ recent instruction-tuned models, under the assumption that more recent models tend to be more reliable.
In practice, the best-performing models show strong precision: while they occasionally introduce false negatives (e.g., treating certain S-CAFs as C-CAFs), they produce virtually no false positives. 
Because false negatives do not compromise soundness, the overall analysis remains conservative and sound.

\section{Related Work}

\subsection{Pointer Analysis}

Recent efforts in pointer analysis have largely aimed to improve the scalability of precise techniques by minimizing redundant computations. Key approaches include applying selective context-sensitivity~\cite{Graphick, Zipper, Scaler}, leveraging demand-driven analysis~\cite{DDPA, SUPA, SUPA_extend}, eliminating propagation redundancies~\cite{PUS, DEA, WP_DP}, and utilizing summary-based~\cite{Calysto, Saturn, pinpoint, Falcon} abstractions.

Graphick~\cite{Graphick} adopts a learning-guided strategy to determine where context sensitivity is most beneficial, outperforming prior techniques such as Zipper~\cite{Zipper} and Scaler~\cite{Scaler} in targeting critical code regions.
PUS~\cite{PUS} constructs a causal graph to identify essential propagation paths, enabling each iteration of Andersen’s algorithm~\cite{Andersen} to focus on just 4\% of the total analysis space.
VSFS~\cite{VSFS} optimizes flow-sensitive pointer analysis by collapsing value-flow nodes into shared versions, significantly reducing computational overhead from redundant updates.
Falcon~\cite{Falcon}, on the other hand, applies summary-based, path-sensitive reasoning to pointer and data dependence analysis, offering an effective balance between accuracy and efficiency.
Ma et al.~\cite{Cutshortcut} observe that Java programs often exhibit merge flows due to features like getters and setters. 
By adapting Andersen’s rules to account for these merge flows, they achieve partial context sensitivity with lower overhead than even context-insensitive approaches.

In addition, the growing adoption of emerging languages like Rust has prompted research~\cite{RUPTA, StackFiltering} into pointer analysis frameworks tailored to their unique language features, and more studies are likely to follow in this direction in the future.

\subsection{Memory Management Function Identification}

Because C requires programmers to manually manage memory (e.g., through \texttt{malloc} and \texttt{free}), real-world software systems often introduce custom memory-management functions, including allocators and deallocators, to simplify maintenance and encapsulate domain-specific memory-handling logic.
To enable effective detection of memory leaks, use-after-free, double-free, and other memory-safety issues, prior research has therefore devoted substantial effort to automatically identifying these custom memory managers.

KMeld~\cite{KMeld} proposes a heuristic strategy that classifies a function as an allocator if it returns a pointer and its callers immediately perform initialization. While this approach is effective for the Linux kernel, its coarse-grained patterns lead to high false-positive rates on other projects.
SinkFinder~\cite{SinkFinder} later employs a form of analogical reasoning: given an example function pair, it searches for other pairs exhibiting similar behavioral patterns. However, because many allocators do not have a clearly corresponding deallocator, and due to the inherent limitations of the NLP models used, SinkFinder still produces considerable false positives and false negatives.
To address these limitations, Raisin~\cite{Raisin} introduces a new hypothesis: allocators generally appear within specific type contexts. 
By extracting function sequences and learning embeddings via NLP models, Raisin identifies rare allocators and deallocators through similarity analysis.
In parallel, Goshawk~\cite{Goshawk} tackles the additional challenge that allocated objects may have structured or hierarchical layouts. Building on allocator/deallocator identification, it incorporates data-flow analysis to remove spurious candidates and reconstruct memory-operation sequences, enabling the detection of nested allocation and deallocation behaviors and improving memory-bug detection effectiveness.

Despite their promise, these approaches are not directly applicable to pointer-analysis frameworks. A central limitation is that they fail to model critical interactions between an allocator and its caller.
For example, storing an allocated object into a field of a caller-provided structure. Missing such interactions leads to unsoundness, constraining their usefulness for sound pointer analysis tasks.

\subsection{LLM-enhanced Program Analysis}

LLMs have been increasingly applied to static analysis, mainly for vulnerability detection, in two ways: enhancing existing tools or directly identifying vulnerabilities.

In the first line, LLift~\cite{LLift} and BugLens uses LLMs to analyze complex vulnerability constraints, reducing false positives in tools like UBITech~\cite{UBITech} and SUTURE~\cite{SUTURE}. LATTE~\cite{Latte} and IRIS~\cite{IRIS} leverage LLMs to identify taint sources and sinks from third-party libraries, improving handling of external code and CodeQL~\cite{codeql} reports. InferROI~\cite{InferROI} detects custom resource management functions to simplify leak detection. GPTScan~\cite{GPTScan} finds risky smart contract operations, aided by static reachability. Lara~\cite{Lara} enhances IoT firmware taint analysis by spotting external input functions, and Artemis~\cite{Artemis} detects injection flaws in PHP libraries via LLM-based input tracing.
The second approach uses LLMs more directly. 
Du et al.\cite{Vul-RAG} proposed Vul-RAG, a retrieval-augmented generation method that matches vulnerable or fixed code snippets from a database to target code. 
Yang et al.\cite{Knighter} introduced KNighter, which generates and reuses checkers from training data to detect similar bugs.

Our work employs simplified prompts to maintain lightweight analysis overhead, in contrast to their approach which requires sophisticated LLM techniques. This design choice aligns with our goal of creating an efficient pre-analysis step that avoids the computational burden of complex LLM interactions.

\section{Conclusion}

We first conduct a preliminary study to demonstrate that accurately identifying custom allocation functions (CAFs) can significantly improve the precision of pointer analysis. Building on this insight, we propose CAFD, an automated framework for detecting CAFs that leverages large language models (LLMs) to enable efficient and lightweight identification. 
The CAFs detected by our approach enhance allocation-based heap modeling, thereby refining pointer analysis results. 
Moreover, this improved pointer analysis facilitates the discovery of additional allocation sources, leads to better indirect call resolution, and enhances memory bug detection.
Nevertheless, achieving further gains in detection accuracy will require the development of more advanced bug detection techniques.

\bibliographystyle{unsrt}
\bibliography{sections/references}

\end{document}